\documentclass[aps,prb,twocolumn,
	groupedaddress,superscriptaddress,
	amsfonts,amssymb,amsmath,floatfix,
	citeautoscript]{revtex4-1}

\usepackage{graphicx}
\usepackage[centering,hmargin=20mm,tmargin=30mm,bmargin=25mm]{geometry}
\usepackage{multirow}
\usepackage{newtxtext}
\usepackage[cmintegrals]{newtxmath}

\usepackage{xcolor}
\usepackage{hyperref}
\hypersetup{colorlinks,
	linkcolor={blue!75!black!80!yellow},
	citecolor={blue!75!black!80!yellow},
	urlcolor={blue!75!black!80!yellow}
}

\makeatletter
\renewcommand\@make@capt@title[2]{%
	\@ifx@empty\float@link{\@firstofone}{\expandafter\href\expandafter{\float@link}}%
	\sffamily{\textbf{#1}}\@caption@fignum@sep#2
}%

\makeatother

\newcommand{\iu}{\mathrm{i}}
\newcommand{\e}{\mathrm{e}}
\newcommand{\dd}{\mathrm{d}}
\newcommand{\unitvec}[1]{\ensuremath{\hat{\mathbf{#1}}}}
\newcommand{\HarvardSEAS}{John A. Paulson School of Engineering and Applied Sciences, Harvard University, Cambridge, MA, USA}
\newcommand{\MITPhy}{Department of Physics, Massachusetts Institute of Technology, Cambridge, MA, USA}

\usepackage[normalem]{ulem}

\begin{document}

\title{\emph{Ab initio} calculation of phonon polaritons in silicon carbide and boron nitride}

\author{Nicholas Rivera$^\perp$}\email{nrivera@seas.harvard.edu}\affiliation{\HarvardSEAS}\affiliation{\MITPhy}
\author{Jennifer Coulter$^\perp$}\affiliation{\HarvardSEAS}
\author{Thomas Christensen}\affiliation{\MITPhy}
\author{Prineha Narang}\email{prineha@seas.harvard.edu}\affiliation{\HarvardSEAS}

\date{\today}

\begin{abstract}
The ability to use photonic quasiparticles to control electromagnetic energy far below the diffraction limit is a defining paradigm in nanophotonics. An important recent development in this field is the measurement and manipulation of extremely confined phonon-polariton modes in polar dielectrics such as silicon carbide and hexagonal boron nitride, which pave the way for nanophotonics and extreme light-matter interactions in the mid-IR to THz frequency range. To further advance this promising field, it is of great interest to predict the optical response of recently discovered and yet-to-be-synthesized polaritonic materials alike. Here we develop a unified framework based on quantum linear response theory to calculate the spatially non-local dielectric function of a polar lattice in arbitrary dimensions. In the case of a three-dimensional bulk material, the spatially local limit of our calculation reproduces standard results for the dielectric response of a polar lattice. Using this framework, we provide \emph{ab initio} calculations of the dielectric permittivity of important bulk polar dielectrics such as silicon carbide and hexagonal boron nitride in good agreement with experiments. From the \emph{ab initio} theory, we are able to develop a microscopic understanding of which phonon modes contribute to each component of the dielectric function, as well as predict features in the dielectric function that are a result of weak TO phonons. This formalism also identifies regime(s) where quantum nonlocal effects may correct the phonon polariton dispersion, extremely relevant in recent atomic-scale experiments which confine electromagnetic fields to the scale of 1~nm.  Finally, our work points the way towards first principles descriptions of the effect of interface phonons, phonon strong coupling, and chiral phonons on the properties of phonon polaritons.
\end{abstract}

\maketitle

Phonon polaritons, quasiparticles of jointly photonic and phononic character, offer great promise for deeply sub-diffractional control of electromagnetic fields at mid-IR and THz frequencies. Phonon polaritons share many features in common with plasmon polaritons in conductors. In recent years, it has been shown that phonon polaritons enable confinement of light to volumes over $10^6$ times smaller than that of a diffraction-limited photon in free-space\cite{caldwell2013low,xu2014mid,caldwell2014sub,dai2014tunable,tomadin2015accessing,yoxall2015direct,li2015hyperbolic,
dai2015subdiffractional,dai2015graphene,caldwell2015low,li2016reversible,Basov:2016,basov2017towards,low2017polaritons,giles2017ultra}. Due to this remarkable confinement and their relatively high lifetimes of around picoseconds, phonon polaritons open new opportunities for vibrational spectroscopy, radiative heat transfer \cite{hillenbrand2002phonon}, and control of dynamics in quantum emitters \cite{kumar2015tunable,rivera2017making,kurman2018control, Peyskens:2018fk}. The core features of phonon polaritons, such as mode shape, confinement, and propagation characteristics, are understood from simple Lorentz oscillator models of the dielectric function, enabling successful theoretical accounts of experimental observations.

Nevertheless, several key questions remain, which are beyond the scope of conventional phenomenological oscillator models and motivate our work. We note a few:
First, what is the range of validity of the otherwise successful Lorentz oscillator model? One mechanism of breakdown concerns the influence of spatial dispersion, i.e nonlocality: recent measurements and (classical) theoretical predictions of extremely confined phonon polaritons, e.g., in nanometer-thin films, underscore the urgency of this question.\cite{Alcaraz-Iranzo:2018uq}

Second, what are the distinguishing properties of phonon polaritons in reduced dimensions, such as in two-dimensional (2D) polar insulators. Can they support sufficiently confined polaritons in analogy to extremely confined plasmonic modes in 2D plasmonic materials such as graphene~\cite{jablan2009plasmonics, koppens2011graphene, fei2012gate, chen2012optical,rivera2016shrinking,lundeberg2017tuning}? Such nanometer-scale confinement could enable non-perturbative light-matter interactions between polaritons and emitters. In general, how do phonon polaritons reflect/inherit the properties of the optical phonons? Finally, what is the influence of non-trivial spatial shaping of phonons on the resulting polariton? Examples of non-trivial spatial shaping include chiral phonons~\cite{zhu2018observation} and mechanical heterostructures~\cite{ratchford2018controlling} such as Moir\'e superlattices.  The ability to design phonon polariton materials by designing the underlying optical phonon modes will be a fruitful source of novel polaritonic nano-materials.

To address these questions an \emph{ab initio} framework to predict the phonon contribution to the dielectric tensor in arbitrary settings is essential. We note that methodologies already exist to calculate the phonon contribution to the local dielectric function in a three-dimensional (3D) bulk, as in Ref.~\citenum{Gonze1997dynamical}, and are implemented in \emph{ab initio} materials physics software packages\cite{abinit1,abinit2,abinit3}. In our work, we aim to extend the validity of these approaches to account for the non-zero lifetime of the phonon modes, the effects of finite temperature, the effects of spatial dispersion on the modes, and the effects of reduced dimensionality. 

Here, we develop a framework based on quantum mechanical linear response theory to analyze phonon polaritons in materials which can address all of these effects. Specifically, we derive a framework for calculating the general, nonlocal dielectric function in polar materials. To unify the treatment of these materials with similar efforts to calculate dielectric properties of plasmonic materials, we highlight parallels with the well-known linear response formalism of free electrons, whose quasiparticles are plasmons rather than phonon polaritons. While the framework and expressions we derive here can account for finite temperature, spatial dispersion and reduced-dimension effects, we confine the applications of the formalism to answer how the underlying optical phonon modes impact the spatially-local dielectric function of a bulk polar material. In particular, we calculate the local dielectric function of bulk silicon carbide (SiC) and bulk hexagonal boron nitride (hBN) from first principles, and show that the results agree well with measured values. We provide a microscopic understanding of the dielectric properties of these two important materials in terms of the relevant transverse optical phonon modes and understand which modes display stronger features in the dielectric function based on \emph{ab initio} calculation of their mode strengths. We find that despite the many transverse optical phonon modes in these materials, only a small number have finite strength and contribute to optical response. 

\section{Linear response of phonons in three dimensions}
\begin{figure*}[t]
\includegraphics[width=16cm]{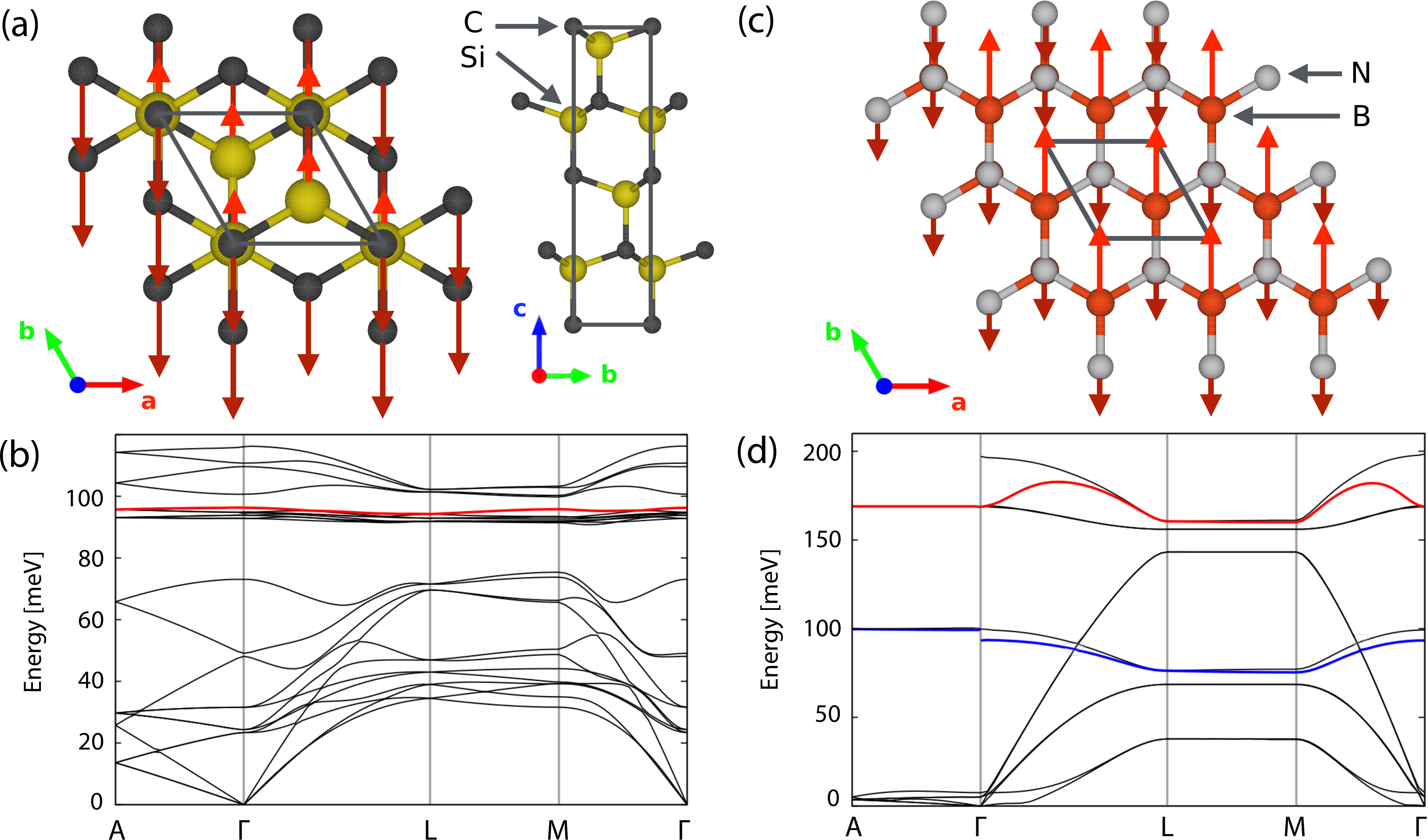}
\caption{
	\textbf{Crystal structure and phonon dispersion of bulk 4H-SiC and hBN.}
	(a)~The hexagonal unit cell structure of SiC allows different optical response in the a-b plane and along the c axis. 
	(b)~The phonon bandstructure of SiC, from which the phonon eigendisplacements associated with the $\Gamma$ point were used to calculate the frequency-dependent dielectric function.
	(c)~Structure (viewed from stacking plane) and 
	(d)~and phonon bandstructure of bulk hBN.}
\label{fig:phonons}
\end{figure*}
A first-principles theory of phonon polaritons can be developed in a manner paralleling the linear response treatment of free electrons, whose resulting quasiparticles are plasmons. Here, rather than free electrons, the polarizable subsystem is the ionic lattice of the crystal. Thus, we develop the linear response theory of an ionic lattice coupled to an electromagnetic field.  The goal of the theory is to find the contribution to the dielectric function from optical phonons. This implicitly constitutes a theory of both bulk and surface phonon \emph{polaritons}. In particular, the bulk phonon polaritons correspond to propagating bulk solutions of Maxwell's equations with the derived permittivity, while surface phonon-polaritons correspond to evanescent solutions at an interface. The latter polaritons, which are of more interest due to their nano-confined properties, exist when the real part of the derived dielectric function is less than zero.

The interaction Hamiltonian of the lattice with the electromagnetic field is:
\begin{equation}\label{eq:interaction}
H_{\text{int}} = -\int \dd^3\mathbf{r}\,\mathbf{E}(\mathbf{r},t)\cdot\mathbf{P}(\mathbf{r},t).
\end{equation}
where $\mathbf{E}$ is the total electric field, and $\mathbf{P}$ is the polarization due to lattice displacements. The polarization can be written as a sum over the dipole moments  $\mathbf{d}(\mathbf{R},t)$ of each unit cell centered at $\{\mathbf{R}\}$:
\begin{equation}\label{eq:dipolesum}
\mathbf{P}(\mathbf{r},t) = \sum\limits_{\mathbf{R}}\mathbf{d}(\mathbf{R},t)\delta(\mathbf{r}-\mathbf{R}),
\end{equation}
This formulation of the polarization implicitly assumes that the ionic displacements are far smaller than the unit cell's extent; this is an exceedingly good approximation, even in a deeply anharmonic regime.
This said, we operate in the harmonic regime.
The dipole moment is given by $\mathbf{d}(\mathbf{R},t) = \sum\limits_{\kappa} \mathbf{Z}_{\kappa}\mathbf{u}(\mathbf{R}_{\kappa} \equiv \mathbf{R}+\mathbf{b}_{\kappa},t)$, where $\mathbf{Z}_{\kappa}$ is the tensor of Born effective charges for atom $\kappa$ in the unit cell at basis vector $\mathbf{b}_{\kappa}$ and $\mathbf{u}(\mathbf{R}_{\kappa},t)$ is the displacement of atom $\kappa$ at basis site $\mathbf{R}_{\kappa}$. The Born charges are formally defined by the relation: $Z_{\kappa,ij} = V\lim_{\mathbf{q}\rightarrow 0}\frac{\partial P_{i}}{\partial u_{\kappa j}}$, with $V$ the volume of the unit cell. In other words, they express the charge dynamically induced by a displacement $\partial u_{\kappa j}$ of all of the atoms $\kappa$ along direction $j$. The displacement of $\kappa$ between unit cells is in phase (i.e., it is at zero wavevector). It can be seen that these charges determine the dipole moment of the lattice that couples to the total electric field, as to lowest order in the atomic displacements, the polarization determined by the displacement of atom $\kappa$ is simply $\frac{1}{V}\mathbf{Z}_{\kappa}\mathbf{u}_{\kappa}$ \cite{Gonze1997dynamical}. We note that these charges are different from the partial charges of the polar atoms, as these dynamical charges take into account the response of electrons to the displacement of lattice sites.  In particular, the rearrangement of electron density due to the change in the ionic potential associated with lattice motion is taken into account. This displacement can be expanded in phonon modes as
\begin{equation}\label{eq:phononfield}
\mathbf{u}(\mathbf{R}_{\kappa},t) = \sum\limits_{\mathbf{q}\sigma}\sqrt{\frac{\hbar}{2M_{\kappa}N\omega_{\mathbf{q}\sigma}}}\left( \e^{\iu\mathbf{q}\cdot\mathbf{R}_{\kappa} - \iu\omega_{\mathbf{q}\sigma}t}\unitvec{e}_{\mathbf{q}\sigma}(\mathbf{b}_{\kappa})a_{\mathbf{q}\sigma} + \text{h.c.} \right),
\end{equation}
where $M_{\kappa}$ is the mass of atom at basis site $\kappa$, $\omega_{q\sigma}$ is the phonon frequency at wavevector $\mathbf{q}$ and branch $\sigma$, $N$ is the number of unit cells, and $a_{\mathbf{q}\sigma}$ is the annihilation operator of a phonon at wavevector $\mathbf{q}$ and branch $\sigma$.
For weak driving fields, the expectation value of the polarization will depend linearly on the total field $\mathbf{E}$, the sum of external and induced fields.
We define the linear-response relation for a bulk crystal in Fourier space as $\langle\mathbf{P}(\mathbf{q},\omega)\rangle = \epsilon_0\boldsymbol{\Pi}(\mathbf{q},\omega)\mathbf{E}(\mathbf{q},\omega)$, with $\boldsymbol{\Pi}(\mathbf{q},\omega)$ defining a polarization-polarization response function, and $\langle \rangle$ denoting the ensemble-average of a microscopic quantity.

To connect this polarization-polarization response function to the dielectric function, we consider Maxwell's equations for the electric field sourced by a polarization density:
\begin{equation}\label{eq:maxwellsourced}
\left(-\mathbf{q}\times\mathbf{q}\times - \boldsymbol{\epsilon}_{\infty}\frac{\omega^2}{c^2} \right)\mathbf{E}(\mathbf{q},\omega) = \omega^2\mu_0\mathbf{P}(\mathbf{q},\omega),
\end{equation}
where $\omega$ is the frequency of light under consideration, $\mathbf{P}$ is the classical polarization at that frequency, and $\boldsymbol{\epsilon}_{\infty}$ is the high-frequency dielectric function due to electronic polarization. By high-frequency, we mean high compared to frequencies associated with phonons but low compared to interband scales in the electronic band structure. Now, we identify the average polarization $\langle\mathbf{P}\rangle$ with the classical polarization $\mathbf{P}$ in Maxwell's equations. This is essentially the random-phase approximation (RPA). We note that an analogous procedure is also employed in the linear-response theory of the longitudinal response of electrons, where instead of the polarization density being averaged, the charge density is averaged. And instead of the solving Maxwell's equations for the field, we solve Laplace's equation for the potential. Using the linear response relation $\mathbf{P} = \epsilon_0\boldsymbol{\Pi}\mathbf{E}$, we have that
\begin{equation}\label{eq:maxwellepsilon}
\left(-\mathbf{q}\times\mathbf{q}\times - \left(\boldsymbol{\epsilon}_{\infty} + \boldsymbol{\Pi} \right)\frac{\omega^2}{c^2}\right)\mathbf{E} = 0,
\end{equation}
leading to the identification of of the dielectric tensor $\boldsymbol{\epsilon} = \boldsymbol{\epsilon}_{\infty} + \boldsymbol{\Pi}$.

Next, we explicate the polarization-polarization response function. From the Kubo formula at finite temperature~\cite{giulianivignale}, we obtain
\begin{equation}\label{eq:polarizationkubo}
\boldsymbol{\Pi}(\mathbf{q},\omega) =  \frac{1}{\epsilon_0 V \mathcal{Z}}\sum\limits_{m,n}\frac{\mathbf{P}_{mn}(\mathbf{q})\otimes\mathbf{P}_{nm}(\mathbf{q})}{\hbar\omega + E_{nm}+\iu\Gamma/2}\left(\e^{-\beta E_m}-\e^{-\beta E_n} \right),
\end{equation}
where $m$ and $n$ refer to a set of phononic eigenstates that span the Fock space, $V$ is a normalization volume, and we have written the polarization-polarization susceptibility in Fourier space. 
Here, $\mathbf{P}_{nm} = \langle n|\mathbf{P}(\mathbf{q})|m\rangle$ defines modal polarization elements of $\mathbf{P}(\mathbf{q})$, the Fourier transform of $\mathbf{P}(\mathbf{r},0)$, $E_{nm} (\equiv  E_n-E_m)$ defines Fock-state energy differences, $\mathcal{Z}$ is the grand canonical partition function, and $\beta\equiv 1/k_{\mathrm{B}}T$ (Boltzmann's constant, $k_{\mathrm{B}}$; temperature, $T$).
Additionally, $\Gamma\rightarrow 0^+$ is a positive infinitesimal, enforcing causality.

Next, we focus our attention on the conceptually important zero-temperature limit, where $\lim_{\beta\rightarrow\infty}\e^{-\beta E}/\mathcal{Z} = \big\{\begin{smallmatrix} 0 \text{ if }E > 0\\ 1\text{ if }E = 0 \end{smallmatrix}$.
Then, considering the $\e^{-\beta E_m}/\mathcal{Z}$ term of Equation~\eqref{eq:polarizationkubo}, $m$ must be a zero-phonon state---and, simultaneously, by the linearity of the displacement in creation and annihilation operators, $n$ must be a one-phonon state. For the $\e^{-\beta E_n}/\mathcal{Z}$ term, the reverse holds true. Thus, the relevant polarization matrix elements are $\mathbf{P}_{10}(\mathbf{q})$. These matrix elements, now denoted $\mathbf{d}(\mathbf{q})$, are simply the Fourier transform  of the site-dependent dipole moments:
\begin{equation}\label{eq:fourierpolarization}
 \mathbf{d}(\mathbf{q}) = \sum\limits_{\mathbf{R}}\mathbf{d}(\mathbf{R})\e^{-\iu\mathbf{q}\cdot\mathbf{R}}.
\end{equation}
Substituting the displacement operator from Equation~\eqref{eq:phononfield}, we find that
\begin{subequations}
\begin{equation}\label{eq:phononstodipole}
\mathbf{d}(\mathbf{q}) = \sum\limits_{\sigma}\sqrt{\frac{\hbar N}{2\omega_{\mathbf{q}\sigma}}} \big(\mathbf{S}_{\mathbf{q}\sigma}a_{\mathbf{q}\sigma} + \text{h.c.}\big),
\end{equation}
with 
\begin{equation}\label{eq:oscillatordefinition}
\mathbf{S}_{\mathbf{q}\sigma} = \sum\limits_{\kappa} \frac{1}{\sqrt{M_{\kappa}}}\mathbf{Z}_{\kappa}\unitvec{e}_{\mathbf{q}\sigma}(\mathbf{b}_{\kappa})\e^{\iu\mathbf{q}\cdot\mathbf{b}_{\kappa}}.
\end{equation}
\end{subequations}

This can be expressed in terms of more standard outputs of \emph{ab initio} materials physics methods that calculate phonon properties. In particular, we can take $\unitvec{e}_{\mathbf{q}\sigma}(\mathbf{b}_{\kappa})$, the eigenvectors of the dynamical matrix, and define eigendisplacements via $\boldsymbol{\eta}_{\mathbf{q}\sigma\kappa}=\unitvec{e}_{\mathbf{q}\sigma}(\mathbf{b}_{\kappa})/\sqrt{M_{\kappa}}$, such that $\mathbf{S}_{\mathbf{q}\sigma} = \sum\limits_{\kappa} \mathbf{Z}_{\kappa}\boldsymbol{\eta}_{\mathbf{q}\sigma\kappa}$.
	
Jointly with Equation~\eqref{eq:polarizationkubo}, this enables an explicit, directly evaluable expression for the dielectric function:

\begin{equation}\label{eq:fullsusceptibility}
\boldsymbol{\Pi}(\mathbf{q},\omega) = \frac{1}{V}\sum\limits_{\sigma}\frac{\mathbf{S}_{\mathbf{q}\sigma}\otimes \mathbf{S}_{\mathbf{q}\sigma}^*}{\omega^2_{\mathbf{q}\sigma}-\omega^2},
\end{equation}
where $V$ is the unit cell volume. 
Rigorously, the phonon frequencies $\omega_{\mathbf{q}\sigma}$ are real. However, in the realistic situation where there is phonon dissipation, typically due to electron-phonon and phonon-phonon scattering, we may make take a ``relaxation-time approximation'' which effectively results in expressing $\omega_{\mathbf{q}\sigma} \rightarrow \omega_{\mathbf{q}\sigma} - \iu\Gamma_{\mathbf{q}\sigma}/2$, where $\Gamma_{\mathbf{q}\sigma}$ is the dissipation rate of the phonon. 

Finally, from Equations~\eqref{eq:maxwellepsilon} and \eqref{eq:fullsusceptibility}, we infer the form of the dielectric function
\begin{equation}\label{eq:pitoepsilon}
\boldsymbol{\epsilon}(\mathbf{q},\omega) = \boldsymbol{\epsilon}_{\infty} + \frac{1}{V}\sum\limits_{\sigma}\frac{\mathbf{S}_{\mathbf{q}\sigma}\otimes \mathbf{S}_{\mathbf{q}\sigma}^*}{\omega^2_{\mathbf{q}\sigma}-\omega^2-\iu\omega\Gamma_{\mathbf{q}\sigma}}.
\end{equation}
This expression is in agreement with earlier theoretical accounts of the phonon-contribution to the dielectric function, as e.g. derived by Born and Huang using an equations of motion approach~\cite{BornHuang:1954}.
We note that the dielectric function in Equation~\eqref{eq:pitoepsilon} explicitly depends on wavevector, i.e. it includes spatial dispersion (nonlocal response). The mathematical manifestation of spatial dispersion however is quite different than that in the analogous case of plasmonic response of electrons. For the phonons, the wavevector dependence is ultimately implicit in the phonon frequencies and the oscillator strengths. For an electronic system, the expression for the spatially dispersive dielectric function would still contain a summation over wavevectors representing non-vertical electronic transitions.

Next, we consider a few important simplifications of this general formulation. Suppose we express the dielectric function in the basis spanned by the principal axes of the crystal, and then consider $\epsilon_{ii}(\omega)$ with $i$ denoting one of those principal directions. Consider a situation in which the oscillator strength of Equation (10) is only large for one particular TO mode. Then the quantity $|\mathbf{S}_{i}|^2$ appearing in $\epsilon_{ii}$ has dimensions of squared charge divided by mass. Defining then an effective charge $Q_{\mathrm{eff}}$ and an effective mass $M_{\mathrm{eff}}$, we may parameterize $|\mathbf{S}_{i}|^2=\frac{Q^2_{\mathrm{eff}}}{M_{\mathrm{eff}}}$. This leads to an expression of the dielectric function as:
\begin{equation}\label{eq:lorentzoscillator}
\epsilon_{ii}(\omega) = \epsilon_{\infty} + \frac{nQ_{\mathrm{eff}}^{2}}{\epsilon_0 M_{\mathrm{eff}}}\frac{1}{\omega^2_{\mathrm{TO}}-\omega^2-\iu\omega\Gamma_{\mathrm{TO}}},
\end{equation}
where $n=N/V$. This coincides precisely with the phenomenological Lorentz oscillator model. That said, the general formulation, Equation~\eqref{eq:pitoepsilon}, incorporates additional physical features, such as the tensorial and site-dependent nature of the Born charges, the effects of a complex unit-cell, and explicitly connects fundamental phonon properties---bandstructure and eigendisplacements---with the infrared dielectric function.

Note that Equation~\eqref{eq:pitoepsilon} has a wavevector dependence which is implicit through the phonon dispersion.
Therefore, it is possible that the phonon polariton resonances red-shift or blue-shift, depending on how the relative position of the phonons at finite-wavevector versus zero wavevector.
For example, in the 4H polytype of SiC (4H-SiC), upon moving along the $\Gamma\text{--}L$ direction of the Brillouin zone, there is a red-shift in the transverse optical phonon which, as per Equation (12), red-shifts the onset of $\epsilon <0$ and correspondingly the onset of surface-confined phonon polaritons. The TO phonon is calculated to be at 764~cm$^{-1}$ at $\Gamma$ (approaching $\Gamma$ along the L direction), and red-shifts: by 2~cm$^{-1}$ at 1/10th of the way along the $\Gamma\text{--}L$ direction, by 5~cm$^{-1}$ at 1/5th of the way along the $\Gamma\text{--}L$ direction, and 13~cm$^{-1}$ halfway along the $\Gamma\text{--}L$ direction.

\begin{figure}[t]
\includegraphics[height=9.75cm]{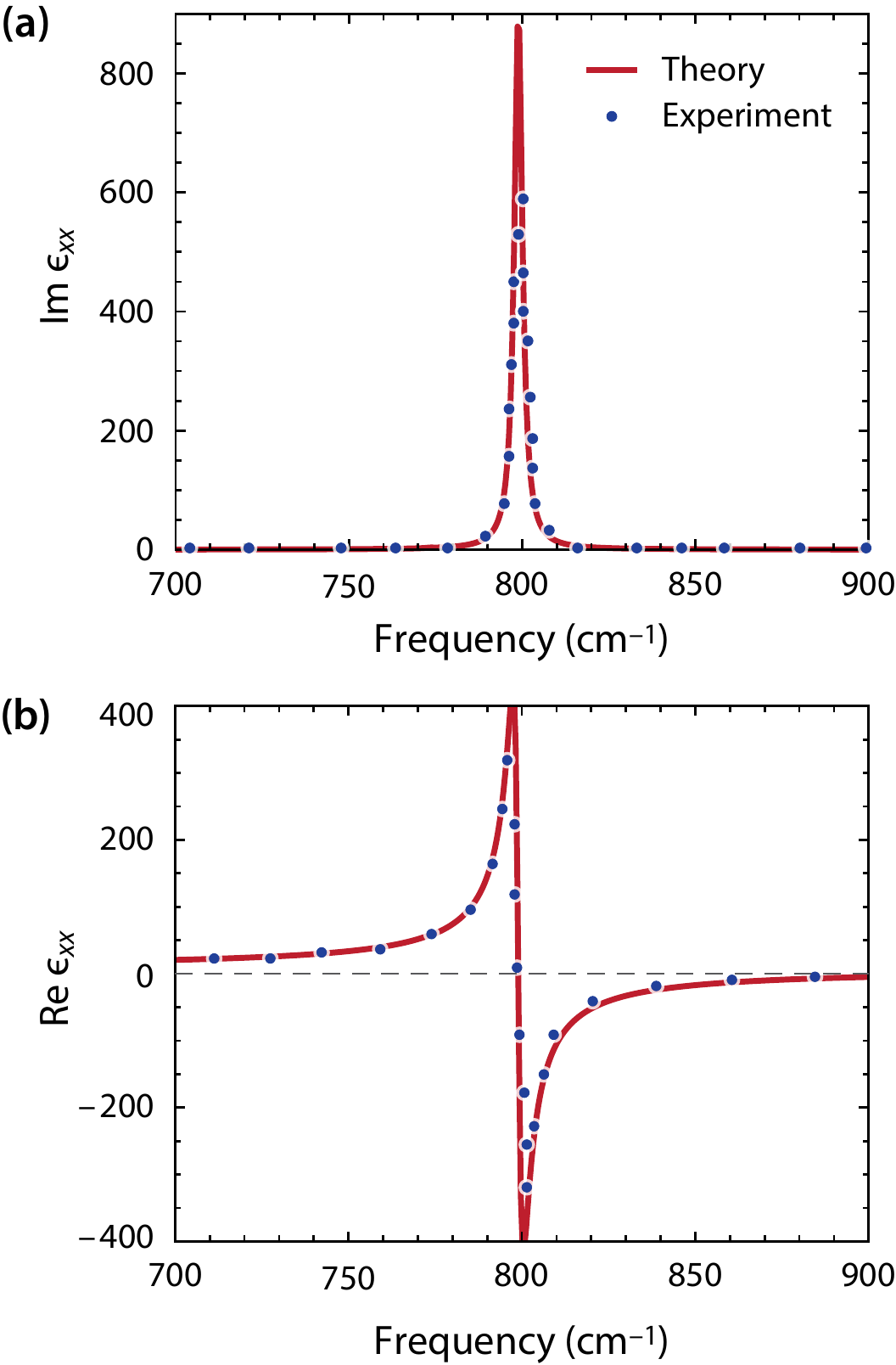}
\caption{%
	\textbf{Comparison of theoretical predictions and experimental values of the infrared dielectric function of bulk 4H-SiC.}
	(a)~Real and (b)~imaginary parts of the frequency-dependent dielectric tensor in the direction orthogonal to the c-axis, calculated for bulk 4H-SiC from Equation~\eqref{eq:pitoepsilon}. Phonon properties and $\epsilon_{\infty}$ calculated from the ABINIT package. 
	Experimental values (Ref.~\citenum{tiwald1999carrier}) are overlaid (blue dots).}
\label{fig:sicepsilon}
\end{figure}

\begin{figure}[t]
\includegraphics[height=9.75cm]{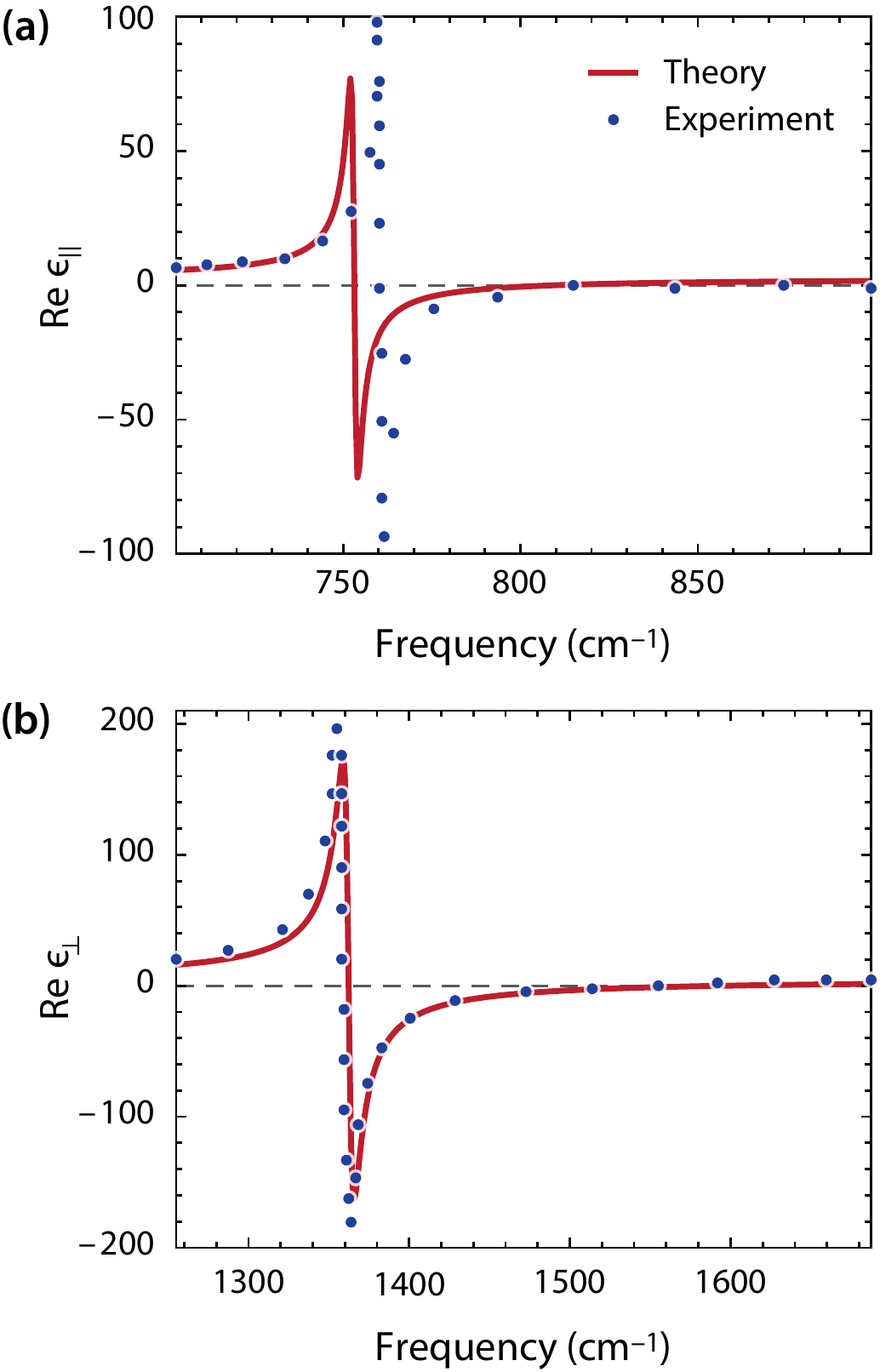}
\caption{%
	\textbf{Comparison of theoretical predictions and experimental values of the infrared dielectric function of bulk hBN.}
	(a) Real part of the frequency-dependent dielectric tensor in a direction orthogonal to the c-axis, calculated for bulk hBN from Equation~\eqref{eq:pitoepsilon}. Phonon properties and $\epsilon_{\infty}$ calculated from the ABINIT package. Experimental values  (Ref.~\citenum{caldwell2014sub}) are overlaid (blue dots). (b) Real part of the frequency-dependent dielectric tensor in the direction along the c-axis.}
\label{fig:hbnepsilon}
\end{figure}

\begin{figure}[t]
\includegraphics[width=1\columnwidth]{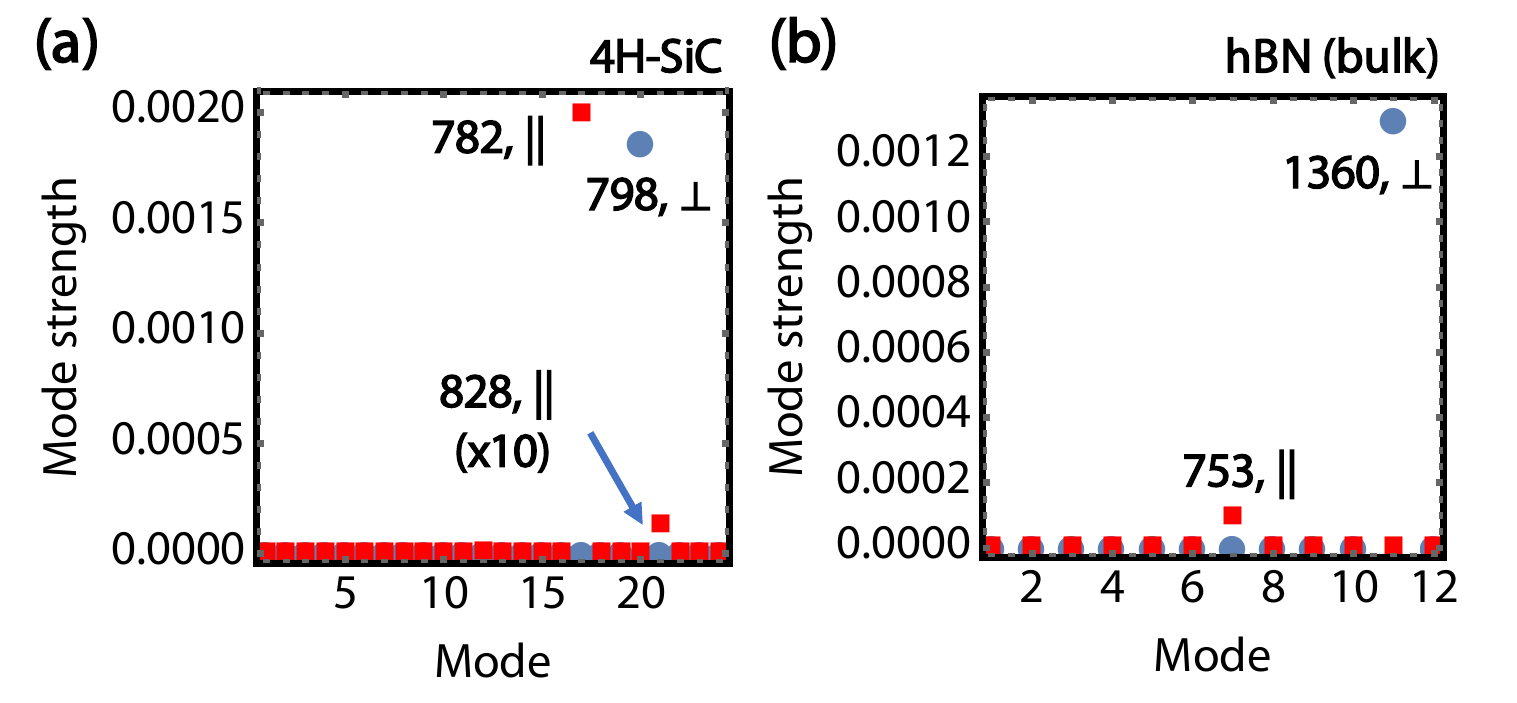}
\caption{%
	\textbf{Mode strengths for transverse optical phonons in silicon carbide and boron nitride.}
	 Diagonal components of the mode strength tensor $ |S_{\sigma,i}|^2$ for transverse optical phonon modes in 4H-SiC (a) and bulk hBN (b). The mode strengths correspond well with the features in the calculated dielectric function and also reveal in the case of 4H-SiC a weak IR-active phonon mode, whose strength nevertheless is too small to be directly apparent in the dielectric function. Red squares denote components along the optic axis, while blue circles denote components perpendicular to it. The weak mode at 828~cm$^{-1}$ in SiC has been multiplied by a factor of 10 in order to be visible.
 	}
\label{fig:oscstrengths}
\end{figure}

In the vast majority of experiments that have been performed on phonon polaritons (e.g., Refs.~\citenum{caldwell2013low,xu2014mid,caldwell2014sub,dai2014tunable,tomadin2015accessing,yoxall2015direct,li2015hyperbolic,dai2015subdiffractional,dai2015graphene,caldwell2015low,li2016reversible,Basov:2016,basov2017towards,low2017polaritons,giles2017ultra})
the wavevector used to probe the polariton is much less than the wavevector scale over which the TO phonon dispersion varies, which is of the order of $\pi/a$ where $a$ is a lattice constant of the polar dielectric.
Even in scattering near-field optical microscope experiments, where one uses an atomic force microscopy tip of radius of about 20~nm to probe the polaritons, the wavevectors accessed would still be quite small compared to the extent of the Brillouin zone.
It would then seem extremely challenging to observe nonlocal behavior of phonon-polaritons.
However, a recent experiment \cite{Alcaraz-Iranzo:2018uq} has shown that plasmons in graphene, by means of a gold mirror, can be confined to dimensions of about 1~nm in the dimension transverse to the graphene sheet.
Additionally, another recent experiment \cite{benz2016single} has leveraged so-called `picocavities', sub-nanometer gaps between a metal nanoparticle and metal film, to access very strong light-matter couplings and extreme variations in fields. These two experiments show a way to get large optically-accessed wavevectors to probe nonlocal behavior in polar dielectrics. In particular, by creating a gap between a polar dielectric film and a metallic nano-antenna where a strongly confined field can be supported, sufficiently high-wavevector modes can be created such that the nonlocal response can be probed.
	
\section{Application to bulk polar dielectric systems}

To demonstrate the utility of this theoretical framework in real materials, we calculate the dielectric function of both 4H-SiC and hBN (see Figure~\ref{fig:phonons}(a,c)). 
Phonon bandstructures for 4H-SiC and hBN were calculated from density functional perturbation
theory (DFPT) as implemented in the ABINIT package\cite{abinit1,abinit2,abinit3,Gonze1997dynamical,Hamann2005metric}, with electric field response included to appropriately capture splitting between longitudinal optical (LO) and transverse optical (TO) phonon modes.
Initial ground state density functional theory (DFT) properties were obtained using norm-conserving Vanderbilt pseudopotentials of the PBEsol parameterization for SiC, and PBE with a D2 (Grimme) van der Waals correction for hBN \cite{PBEsol,PBE,ONCV_PPs,pseudodojo,grimme}.
We calculated the phonon modes, their associated eigenfrequencies (Figure~\ref{fig:phonons}(b)) and eigendisplacements at the $\Gamma$ point to construct the local (i.e. $q\rightarrow 0$) dielectric function from Equation~\eqref{eq:pitoepsilon}, using experimental values for the dissipation rates $\Gamma$, taken from \cite{tiwald1999carrier} for SiC and \cite{caldwell2014sub} for hBN. For each case, the dissipation rate taken in each plot is that corresponding to the TO phonon mode with the largest oscillator strength in the frequency windows plotted.

Additionally, for SiC a scissor shift of 1.01~eV was applied to shift the calculated band gap of 2.26~eV to match the experimental gap of $\sim\text{3.27~eV}$ \cite{band_gap_expt1,band_gap_expt2}.
This is necessary to correct for the well-known underestimation of the band gap in ground state density functional theory calculations and the corresponding effects to the electronic contribution to the dielectric tensor, as well as the phonon frequencies. While we have presently taken both the SiC bandgap and the SiC optical phonon lifetime from experiments, these quantities may also be obtained \emph{ab initio} by calculating lifetimes associated with electron-phonon scattering \cite{sundararaman2014theoretical,brown2015nonradiative,brown2016ab,narang2017effects} and three-phonon decay processes \cite{srivastava1990physics}. In materials like hBN where excitonic effects play a significant role, the optical properties of the system are not determined solely by the energy associated with band to band recombination, and instead must also account for the exciton binding energy\cite{robert2016excitonic}. 
Therefore, for the calculation of the dielectric function of hBN, we consider the optical gap, $E_{\mathrm{opt}}$ = $E_{\mathrm{g}} - E_{\mathrm{b}}$, with the band gap, $E_{\mathrm{g}} = \text{5.2~eV}$, from experiment\cite{levinshtein2001properties} and the binding energy, $E_{\mathrm{b}} = \text{0.67~eV}$, from recent \emph{ab initio} calculations\cite{attaccalite2018two}, so that $E_{\mathrm{opt}}$ is approximately 4.5~eV. As the bandgap we compute from DFT is 4.48~eV, we did not require a correction to the gap in hBN.

With these prescriptions, the high-frequency dielectric tensor was determined to be $\epsilon_{\infty,\parallel} = 7.0$ and $\epsilon_{\infty,\perp} = 7.3$ for SiC; $\epsilon_{\infty,\parallel} = 4.6$ and $\epsilon_{\infty,\perp} = 2.6$ for hBN. The values of $\boldsymbol{\epsilon}_{\infty}$ are incorporated in the evaluation of---and consequently impact---the phonon bandstructure.
If the material's band gap is less well-known than e.g. SiC's, or if a strict \emph{ab initio} outlook is desired, the above approach can be amended by a GW calculation. Further, if excitonic effects contribute significantly to the optical properties of a chosen material, a Bethe--Salpeter equation prediction might be used to further improve predictive accuracy within the same framework presented here.
	
The resulting frequency-dependent dielectric tensor, compared to experimental measurements~\cite{tiwald1999carrier,caldwell2014sub}, is shown in Figures~\ref{fig:sicepsilon} and \ref{fig:hbnepsilon}. 
The agreement between the theoretical calculation and the experimental results is excellent for SiC, as well as for hBN in the direction perpendicular to the optical axis. For the component of the hBN dielectric function parallel to the optical axis, a minor discrepancy develops due to a $\sim$ 5\% deviation between calculated and measured TO phonon frequencies. 
In Figure \ref{fig:oscstrengths}, we plot the mode strengths for the different phonon modes contributing to the dielectric response of 4H-SiC and hBN. For SiC, we see a strong mode at 798~cm$^{-1}$, corresponding to the feature found in the permittivity components perpendicular to the $c$-axis. Additionally, there is a strong mode at 782~cm$^{-1}$, which will manifest itself as a strong response in the component of the permittivity parallel to the $c$-axis, corresponding well to experimental measurements\cite{tiwald1999carrier}. Moreover, there appears to be a weak mode at 828~cm$^{-1}$ whose mode strength is 100 times smaller than that of the strong modes. It is interesting that despite the great multiplicity of phonon modes, only a small number lead to the optical response. It will be an interesting area of future study to engineer systems where many nearby TO phonons contribute to the overall dielectric and polaritonic response. In the case of hBN, we see the two modes at 1360~cm$^{-1}$ and 753~cm$^{-1}$ corresponding to features in the dielectric response perpendicular and parallel to the $c$-axis, as expected. Interestingly here, the oscillator strength of the low frequency mode is much smaller than that of the high-frequency mode, leading to a substantially weaker dielectric response. It is partially compensated in the dielectric function however by a substantially longer lifetime for the lower frequency mode, measured to be about 2~cm$^{-1}$ for the low frequency mode and 7~cm$^{-1}$ for the high frequency mode, as in Ref.~\citenum{caldwell2014sub}.

\section{ Outlook}

In summary, we have provided a theoretical framework based on linear response theory to calculate the phonon contribution to the dielectric function from first principles. Notably, we go beyond oversimplifications of the Lorentz oscillator model in which the Born charges are treated as a single scalar quantity and can treat the influence and interplay of many phonon modes that contribute to the dielectric function. We corroborated our approach through density functional theoretic calculations with an accurate prediction of the dielectric function of hBN and SiC, two of the most important phonon-polaritonic materials.

This framework is versatile, allowing us to use first principles calculations to get the dielectric function and predict how nonlocality enters the dielectric function. It also enables an approach to questions regarding the impact of reduced dimensionality in phonon polaritonics; questions that we are presently pursuing. We applied the formalism to the calculation of the local permittivity of silicon carbide and hexagonal boron nitride, two phononic materials of great interest in nanophotonics. Besides accurate prediction of these dielectric properties, we are able to develop a microscopic understanding of the contribution of the many different optical phonon modes to the observed dielectric function of these materials. In particular, we find that while each of these materials has many optical phonon modes (18 in silicon carbide and 9 in hexagonal boron nitride), all but a very small number of modes have oscillator strength that contribute to the dielectric function. In addition, we were also able to find with this microscopic approach phonon modes with weak oscillator strength that may contribute features to the dielectric function.

In future work, besides explicit \emph{ab initio} calculations of the impact of nonlocality and an application of the framework to phonon polaritons in systems of reduced dimensionality, there are a number of interesting directions  that can be addressed by the framework discussed here. One such direction would be to find a system where the optical phonons are drastically different from 3D to 2D. Perhaps it is possible that there are some materials in which the 2D optical phonons experience lower losses due to a reduced scattering phase space. The formalism we provide here may also be extended to understand phonon-polaritons in other more atypical reduced-dimensional settings, such as zero-dimensional settings in single emitters, i.e., `molecular phonon polaritons', in analogy to recent work on `molecular plasmons' \cite{manjavacas2013tunable,lauchner2015molecular}. Another such question is whether optical interface-phonons between adjacent heterogeneous materials could host novel types of phonon polaritons due to strong coupling of the optical phonons between layers. In that case, it would be relevant to evaluate how this strong coupling manifests itself in the infrared dielectric function, and ultimately the confinement and propagation of the phonon polaritons.

\section{Acknowledgements}
The authors thank Joshua Caldwell (Vanderbilt University), Dominik Maximilian Juraschek (ETH Zurich) and Johannes Flick (Harvard University) for helpful discussions. This research used resources of the National Energy Research Scientific Computing Center, a DOE Office of Science User Facility supported by the Office of Science of the U.S. Department of Energy under Contract No. DE-AC02-05CH11231 and computational facilities at Harvard Research Computing. N. R. and J. C. recognize the support of the DOE Computational Science Graduate Fellowship (CSGF) Number DE-FG02-97ER25308. T. C. acknowledges support from the Danish Council for Independent Research (Grant No.\ DFF--6108-00667). The authors acknowledge funding and support from the STC Center for Integrated Quantum Materials NSF grant number DMR-1231319.

\bibliographystyle{apsrev4-1}
\bibliography{references}

\begin{thebibliography}{56}%
\makeatletter
\providecommand \@ifxundefined [1]{%
 \@ifx{#1\undefined}
}%
\providecommand \@ifnum [1]{%
 \ifnum #1\expandafter \@firstoftwo
 \else \expandafter \@secondoftwo
 \fi
}%
\providecommand \@ifx [1]{%
 \ifx #1\expandafter \@firstoftwo
 \else \expandafter \@secondoftwo
 \fi
}%
\providecommand \natexlab [1]{#1}%
\providecommand \enquote  [1]{``#1''}%
\providecommand \bibnamefont  [1]{#1}%
\providecommand \bibfnamefont [1]{#1}%
\providecommand \citenamefont [1]{#1}%
\providecommand \href@noop [0]{\@secondoftwo}%
\providecommand \href [0]{\begingroup \@sanitize@url \@href}%
\providecommand \@href[1]{\@@startlink{#1}\@@href}%
\providecommand \@@href[1]{\endgroup#1\@@endlink}%
\providecommand \@sanitize@url [0]{\catcode `\\12\catcode `\$12\catcode
  `\&12\catcode `\#12\catcode `\^12\catcode `\_12\catcode `\%12\relax}%
\providecommand \@@startlink[1]{}%
\providecommand \@@endlink[0]{}%
\providecommand \url  [0]{\begingroup\@sanitize@url \@url }%
\providecommand \@url [1]{\endgroup\@href {#1}{\urlprefix }}%
\providecommand \urlprefix  [0]{URL }%
\providecommand \Eprint [0]{\href }%
\providecommand \doibase [0]{http://dx.doi.org/}%
\providecommand \selectlanguage [0]{\@gobble}%
\providecommand \bibinfo  [0]{\@secondoftwo}%
\providecommand \bibfield  [0]{\@secondoftwo}%
\providecommand \translation [1]{[#1]}%
\providecommand \BibitemOpen [0]{}%
\providecommand \bibitemStop [0]{}%
\providecommand \bibitemNoStop [0]{.\EOS\space}%
\providecommand \EOS [0]{\spacefactor3000\relax}%
\providecommand \BibitemShut  [1]{\csname bibitem#1\endcsname}%
\let\auto@bib@innerbib\@empty
%</preamble>
\bibitem [{\citenamefont {Caldwell}\ \emph {et~al.}(2013)\citenamefont
  {Caldwell} \emph {et~al.}}]{caldwell2013low}%
  \BibitemOpen
  \bibfield  {author} {\bibinfo {author} {\bibfnamefont {J.~D.}\ \bibnamefont
  {Caldwell}} \emph {et~al.},\ }\href {\doibase 10.1021/nl401590g} {\bibfield
  {journal} {\bibinfo  {journal} {Nano Lett.}\ }\textbf {\bibinfo {volume}
  {13}},\ \bibinfo {pages} {3690} (\bibinfo {year} {2013})}\BibitemShut
  {NoStop}%
\bibitem [{\citenamefont {Xu}\ \emph {et~al.}(2014)\citenamefont {Xu},
  \citenamefont {Jiang}, \citenamefont {Gilburd}, \citenamefont {Rensing},
  \citenamefont {Burch}, \citenamefont {Zhi}, \citenamefont {Bando},
  \citenamefont {Golberg},\ and\ \citenamefont {Walker}}]{xu2014mid}%
  \BibitemOpen
  \bibfield  {author} {\bibinfo {author} {\bibfnamefont {X.~G.}\ \bibnamefont
  {Xu}}, \bibinfo {author} {\bibfnamefont {J.-H.}\ \bibnamefont {Jiang}},
  \bibinfo {author} {\bibfnamefont {L.}~\bibnamefont {Gilburd}}, \bibinfo
  {author} {\bibfnamefont {R.~G.}\ \bibnamefont {Rensing}}, \bibinfo {author}
  {\bibfnamefont {K.~S.}\ \bibnamefont {Burch}}, \bibinfo {author}
  {\bibfnamefont {C.}~\bibnamefont {Zhi}}, \bibinfo {author} {\bibfnamefont
  {Y.}~\bibnamefont {Bando}}, \bibinfo {author} {\bibfnamefont
  {D.}~\bibnamefont {Golberg}}, \ and\ \bibinfo {author} {\bibfnamefont
  {G.~C.}\ \bibnamefont {Walker}},\ }\href {\doibase 10.1021/nn504093g}
  {\bibfield  {journal} {\bibinfo  {journal} {ACS Nano}\ }\textbf {\bibinfo
  {volume} {8}},\ \bibinfo {pages} {11305} (\bibinfo {year}
  {2014})}\BibitemShut {NoStop}%
\bibitem [{\citenamefont {Caldwell}\ \emph {et~al.}(2014)\citenamefont
  {Caldwell} \emph {et~al.}}]{caldwell2014sub}%
  \BibitemOpen
  \bibfield  {author} {\bibinfo {author} {\bibfnamefont {J.~D.}\ \bibnamefont
  {Caldwell}} \emph {et~al.},\ }\href {\doibase 10.1038/ncomms6221} {\bibfield
  {journal} {\bibinfo  {journal} {Nat. Commun.}\ }\textbf {\bibinfo {volume}
  {5}},\ \bibinfo {pages} {5221} (\bibinfo {year} {2014})}\BibitemShut
  {NoStop}%
\bibitem [{\citenamefont {Dai}\ \emph {et~al.}(2014)\citenamefont {Dai} \emph
  {et~al.}}]{dai2014tunable}%
  \BibitemOpen
  \bibfield  {author} {\bibinfo {author} {\bibfnamefont {S.}~\bibnamefont
  {Dai}} \emph {et~al.},\ }\href {\doibase 10.1126/science.1246833} {\bibfield
  {journal} {\bibinfo  {journal} {Science}\ }\textbf {\bibinfo {volume}
  {343}},\ \bibinfo {pages} {1125} (\bibinfo {year} {2014})}\BibitemShut
  {NoStop}%
\bibitem [{\citenamefont {Tomadin}\ \emph {et~al.}(2015)\citenamefont
  {Tomadin}, \citenamefont {Principi}, \citenamefont {Song}, \citenamefont
  {Levitov},\ and\ \citenamefont {Polini}}]{tomadin2015accessing}%
  \BibitemOpen
  \bibfield  {author} {\bibinfo {author} {\bibfnamefont {A.}~\bibnamefont
  {Tomadin}}, \bibinfo {author} {\bibfnamefont {A.}~\bibnamefont {Principi}},
  \bibinfo {author} {\bibfnamefont {J.~C.}\ \bibnamefont {Song}}, \bibinfo
  {author} {\bibfnamefont {L.~S.}\ \bibnamefont {Levitov}}, \ and\ \bibinfo
  {author} {\bibfnamefont {M.}~\bibnamefont {Polini}},\ }\href {\doibase
  10.1103/PhysRevLett.115.087401} {\bibfield  {journal} {\bibinfo  {journal}
  {Phys. Rev. Lett.}\ }\textbf {\bibinfo {volume} {115}},\ \bibinfo {pages}
  {087401} (\bibinfo {year} {2015})}\BibitemShut {NoStop}%
\bibitem [{\citenamefont {Yoxall}\ \emph {et~al.}(2015)\citenamefont {Yoxall},
  \citenamefont {Schnell}, \citenamefont {Nikitin}, \citenamefont {Txoperena},
  \citenamefont {Woessner}, \citenamefont {Lundeberg}, \citenamefont
  {Casanova}, \citenamefont {Hueso}, \citenamefont {Koppens},\ and\
  \citenamefont {Hillenbrand}}]{yoxall2015direct}%
  \BibitemOpen
  \bibfield  {author} {\bibinfo {author} {\bibfnamefont {E.}~\bibnamefont
  {Yoxall}}, \bibinfo {author} {\bibfnamefont {M.}~\bibnamefont {Schnell}},
  \bibinfo {author} {\bibfnamefont {A.~Y.}\ \bibnamefont {Nikitin}}, \bibinfo
  {author} {\bibfnamefont {O.}~\bibnamefont {Txoperena}}, \bibinfo {author}
  {\bibfnamefont {A.}~\bibnamefont {Woessner}}, \bibinfo {author}
  {\bibfnamefont {M.~B.}\ \bibnamefont {Lundeberg}}, \bibinfo {author}
  {\bibfnamefont {F.}~\bibnamefont {Casanova}}, \bibinfo {author}
  {\bibfnamefont {L.~E.}\ \bibnamefont {Hueso}}, \bibinfo {author}
  {\bibfnamefont {F.~H.}\ \bibnamefont {Koppens}}, \ and\ \bibinfo {author}
  {\bibfnamefont {R.}~\bibnamefont {Hillenbrand}},\ }\href {\doibase
  10.1038/nphoton.2015.166} {\bibfield  {journal} {\bibinfo  {journal} {Nat.
  Photonics}\ }\textbf {\bibinfo {volume} {9}},\ \bibinfo {pages} {674}
  (\bibinfo {year} {2015})}\BibitemShut {NoStop}%
\bibitem [{\citenamefont {Li}\ \emph {et~al.}(2015)\citenamefont {Li},
  \citenamefont {Lewin}, \citenamefont {Kretinin}, \citenamefont {Caldwell},
  \citenamefont {Novoselov}, \citenamefont {Taniguchi}, \citenamefont
  {Watanabe}, \citenamefont {Gaussmann},\ and\ \citenamefont
  {Taubner}}]{li2015hyperbolic}%
  \BibitemOpen
  \bibfield  {author} {\bibinfo {author} {\bibfnamefont {P.}~\bibnamefont
  {Li}}, \bibinfo {author} {\bibfnamefont {M.}~\bibnamefont {Lewin}}, \bibinfo
  {author} {\bibfnamefont {A.~V.}\ \bibnamefont {Kretinin}}, \bibinfo {author}
  {\bibfnamefont {J.~D.}\ \bibnamefont {Caldwell}}, \bibinfo {author}
  {\bibfnamefont {K.~S.}\ \bibnamefont {Novoselov}}, \bibinfo {author}
  {\bibfnamefont {T.}~\bibnamefont {Taniguchi}}, \bibinfo {author}
  {\bibfnamefont {K.}~\bibnamefont {Watanabe}}, \bibinfo {author}
  {\bibfnamefont {F.}~\bibnamefont {Gaussmann}}, \ and\ \bibinfo {author}
  {\bibfnamefont {T.}~\bibnamefont {Taubner}},\ }\href {\doibase
  10.1038/ncomms8507} {\bibfield  {journal} {\bibinfo  {journal} {Nat.
  Commun.}\ }\textbf {\bibinfo {volume} {6}},\ \bibinfo {pages} {7507}
  (\bibinfo {year} {2015})}\BibitemShut {NoStop}%
\bibitem [{\citenamefont {Dai}\ \emph {et~al.}(2015{\natexlab{a}})\citenamefont
  {Dai} \emph {et~al.}}]{dai2015subdiffractional}%
  \BibitemOpen
  \bibfield  {author} {\bibinfo {author} {\bibfnamefont {S.}~\bibnamefont
  {Dai}} \emph {et~al.},\ }\href {\doibase 10.1038/ncomms7963} {\bibfield
  {journal} {\bibinfo  {journal} {Nat. Commun.}\ }\textbf {\bibinfo {volume}
  {6}},\ \bibinfo {pages} {6963} (\bibinfo {year}
  {2015}{\natexlab{a}})}\BibitemShut {NoStop}%
\bibitem [{\citenamefont {Dai}\ \emph {et~al.}(2015{\natexlab{b}})\citenamefont
  {Dai} \emph {et~al.}}]{dai2015graphene}%
  \BibitemOpen
  \bibfield  {author} {\bibinfo {author} {\bibfnamefont {S.}~\bibnamefont
  {Dai}} \emph {et~al.},\ }\href {\doibase 10.1038/nnano.2015.131} {\bibfield
  {journal} {\bibinfo  {journal} {Nat. Nanotechnol.}\ }\textbf {\bibinfo
  {volume} {10}},\ \bibinfo {pages} {682} (\bibinfo {year}
  {2015}{\natexlab{b}})}\BibitemShut {NoStop}%
\bibitem [{\citenamefont {Caldwell}\ \emph {et~al.}(2015)\citenamefont
  {Caldwell}, \citenamefont {Lindsay}, \citenamefont {Giannini}, \citenamefont
  {Vurgaftman}, \citenamefont {Reinecke}, \citenamefont {Maier},\ and\
  \citenamefont {Glembocki}}]{caldwell2015low}%
  \BibitemOpen
  \bibfield  {author} {\bibinfo {author} {\bibfnamefont {J.~D.}\ \bibnamefont
  {Caldwell}}, \bibinfo {author} {\bibfnamefont {L.}~\bibnamefont {Lindsay}},
  \bibinfo {author} {\bibfnamefont {V.}~\bibnamefont {Giannini}}, \bibinfo
  {author} {\bibfnamefont {I.}~\bibnamefont {Vurgaftman}}, \bibinfo {author}
  {\bibfnamefont {T.~L.}\ \bibnamefont {Reinecke}}, \bibinfo {author}
  {\bibfnamefont {S.~A.}\ \bibnamefont {Maier}}, \ and\ \bibinfo {author}
  {\bibfnamefont {O.~J.}\ \bibnamefont {Glembocki}},\ }\href {\doibase
  10.1515/nanoph-2014-0003} {\bibfield  {journal} {\bibinfo  {journal}
  {Nanophotonics}\ }\textbf {\bibinfo {volume} {4}},\ \bibinfo {pages} {44}
  (\bibinfo {year} {2015})}\BibitemShut {NoStop}%
\bibitem [{\citenamefont {Li}\ \emph {et~al.}(2016)\citenamefont {Li},
  \citenamefont {Yang}, \citenamefont {Ma{\ss}}, \citenamefont {Hanss},
  \citenamefont {Lewin}, \citenamefont {Michel}, \citenamefont {Wuttig},\ and\
  \citenamefont {Taubner}}]{li2016reversible}%
  \BibitemOpen
  \bibfield  {author} {\bibinfo {author} {\bibfnamefont {P.}~\bibnamefont
  {Li}}, \bibinfo {author} {\bibfnamefont {X.}~\bibnamefont {Yang}}, \bibinfo
  {author} {\bibfnamefont {T.~W.}\ \bibnamefont {Ma{\ss}}}, \bibinfo {author}
  {\bibfnamefont {J.}~\bibnamefont {Hanss}}, \bibinfo {author} {\bibfnamefont
  {M.}~\bibnamefont {Lewin}}, \bibinfo {author} {\bibfnamefont {A.-K.~U.}\
  \bibnamefont {Michel}}, \bibinfo {author} {\bibfnamefont {M.}~\bibnamefont
  {Wuttig}}, \ and\ \bibinfo {author} {\bibfnamefont {T.}~\bibnamefont
  {Taubner}},\ }\href {\doibase 10.1038/nmat4649} {\bibfield  {journal}
  {\bibinfo  {journal} {Nat. Mater.}\ }\textbf {\bibinfo {volume} {15}},\
  \bibinfo {pages} {870} (\bibinfo {year} {2016})}\BibitemShut {NoStop}%
\bibitem [{\citenamefont {Basov}\ \emph {et~al.}(2016)\citenamefont {Basov},
  \citenamefont {Fogler},\ and\ \citenamefont {Garc\'{i}a~de
  Abajo}}]{Basov:2016}%
  \BibitemOpen
  \bibfield  {author} {\bibinfo {author} {\bibfnamefont {D.~N.}\ \bibnamefont
  {Basov}}, \bibinfo {author} {\bibfnamefont {M.~M.}\ \bibnamefont {Fogler}}, \
  and\ \bibinfo {author} {\bibfnamefont {F.~J.}\ \bibnamefont {Garc\'{i}a~de
  Abajo}},\ }\href {\doibase 10.1126/science.aag1992} {\bibfield  {journal}
  {\bibinfo  {journal} {Science}\ }\textbf {\bibinfo {volume} {354}},\ \bibinfo
  {pages} {aag1992} (\bibinfo {year} {2016})}\BibitemShut {NoStop}%
\bibitem [{\citenamefont {Basov}\ \emph {et~al.}(2017)\citenamefont {Basov},
  \citenamefont {Averitt},\ and\ \citenamefont {Hsieh}}]{basov2017towards}%
  \BibitemOpen
  \bibfield  {author} {\bibinfo {author} {\bibfnamefont {D.}~\bibnamefont
  {Basov}}, \bibinfo {author} {\bibfnamefont {R.}~\bibnamefont {Averitt}}, \
  and\ \bibinfo {author} {\bibfnamefont {D.}~\bibnamefont {Hsieh}},\ }\href
  {\doibase 10.1038/nmat5017} {\bibfield  {journal} {\bibinfo  {journal} {Nat.
  Mater.}\ }\textbf {\bibinfo {volume} {16}},\ \bibinfo {pages} {1077}
  (\bibinfo {year} {2017})}\BibitemShut {NoStop}%
\bibitem [{\citenamefont {Low}\ \emph {et~al.}(2017)\citenamefont {Low},
  \citenamefont {Chaves}, \citenamefont {Caldwell}, \citenamefont {Kumar},
  \citenamefont {Fang}, \citenamefont {Avouris}, \citenamefont {Heinz},
  \citenamefont {Guinea}, \citenamefont {Martin-Moreno},\ and\ \citenamefont
  {Koppens}}]{low2017polaritons}%
  \BibitemOpen
  \bibfield  {author} {\bibinfo {author} {\bibfnamefont {T.}~\bibnamefont
  {Low}}, \bibinfo {author} {\bibfnamefont {A.}~\bibnamefont {Chaves}},
  \bibinfo {author} {\bibfnamefont {J.~D.}\ \bibnamefont {Caldwell}}, \bibinfo
  {author} {\bibfnamefont {A.}~\bibnamefont {Kumar}}, \bibinfo {author}
  {\bibfnamefont {N.~X.}\ \bibnamefont {Fang}}, \bibinfo {author}
  {\bibfnamefont {P.}~\bibnamefont {Avouris}}, \bibinfo {author} {\bibfnamefont
  {T.~F.}\ \bibnamefont {Heinz}}, \bibinfo {author} {\bibfnamefont
  {F.}~\bibnamefont {Guinea}}, \bibinfo {author} {\bibfnamefont
  {L.}~\bibnamefont {Martin-Moreno}}, \ and\ \bibinfo {author} {\bibfnamefont
  {F.}~\bibnamefont {Koppens}},\ }\href {\doibase 10.1038/nmat4792} {\bibfield
  {journal} {\bibinfo  {journal} {Nat. Mater.}\ }\textbf {\bibinfo {volume}
  {16}},\ \bibinfo {pages} {182} (\bibinfo {year} {2017})}\BibitemShut
  {NoStop}%
\bibitem [{\citenamefont {Giles}\ \emph {et~al.}(2017)\citenamefont {Giles}
  \emph {et~al.}}]{giles2017ultra}%
  \BibitemOpen
  \bibfield  {author} {\bibinfo {author} {\bibfnamefont {A.~J.}\ \bibnamefont
  {Giles}} \emph {et~al.},\ }\href {https://arxiv.org/abs/1705.05971}
  {\bibfield  {journal} {\bibinfo  {journal} {arXiv:1705.05971}\ } (\bibinfo
  {year} {2017})}\BibitemShut {NoStop}%
\bibitem [{\citenamefont {Hillenbrand}\ \emph {et~al.}(2002)\citenamefont
  {Hillenbrand}, \citenamefont {Taubner},\ and\ \citenamefont
  {Keilmann}}]{hillenbrand2002phonon}%
  \BibitemOpen
  \bibfield  {author} {\bibinfo {author} {\bibfnamefont {R.}~\bibnamefont
  {Hillenbrand}}, \bibinfo {author} {\bibfnamefont {T.}~\bibnamefont
  {Taubner}}, \ and\ \bibinfo {author} {\bibfnamefont {F.}~\bibnamefont
  {Keilmann}},\ }\href {\doibase 10.1038/nature00899} {\bibfield  {journal}
  {\bibinfo  {journal} {Nature}\ }\textbf {\bibinfo {volume} {418}},\ \bibinfo
  {pages} {159} (\bibinfo {year} {2002})}\BibitemShut {NoStop}%
\bibitem [{\citenamefont {Kumar}\ \emph {et~al.}(2015)\citenamefont {Kumar},
  \citenamefont {Low}, \citenamefont {Fung}, \citenamefont {Avouris},\ and\
  \citenamefont {Fang}}]{kumar2015tunable}%
  \BibitemOpen
  \bibfield  {author} {\bibinfo {author} {\bibfnamefont {A.}~\bibnamefont
  {Kumar}}, \bibinfo {author} {\bibfnamefont {T.}~\bibnamefont {Low}}, \bibinfo
  {author} {\bibfnamefont {K.~H.}\ \bibnamefont {Fung}}, \bibinfo {author}
  {\bibfnamefont {P.}~\bibnamefont {Avouris}}, \ and\ \bibinfo {author}
  {\bibfnamefont {N.~X.}\ \bibnamefont {Fang}},\ }\href {\doibase
  10.1021/acs.nanolett.5b01191} {\bibfield  {journal} {\bibinfo  {journal}
  {Nano Lett.}\ }\textbf {\bibinfo {volume} {15}},\ \bibinfo {pages} {3172}
  (\bibinfo {year} {2015})}\BibitemShut {NoStop}%
\bibitem [{\citenamefont {Rivera}\ \emph {et~al.}(2017)\citenamefont {Rivera},
  \citenamefont {Rosolen}, \citenamefont {Joannopoulos}, \citenamefont
  {Kaminer},\ and\ \citenamefont {Solja{\v{c}}i{\'c}}}]{rivera2017making}%
  \BibitemOpen
  \bibfield  {author} {\bibinfo {author} {\bibfnamefont {N.}~\bibnamefont
  {Rivera}}, \bibinfo {author} {\bibfnamefont {G.}~\bibnamefont {Rosolen}},
  \bibinfo {author} {\bibfnamefont {J.~D.}\ \bibnamefont {Joannopoulos}},
  \bibinfo {author} {\bibfnamefont {I.}~\bibnamefont {Kaminer}}, \ and\
  \bibinfo {author} {\bibfnamefont {M.}~\bibnamefont {Solja{\v{c}}i{\'c}}},\
  }\href {\doibase 10.1073/pnas.1713538114} {\bibfield  {journal} {\bibinfo
  {journal} {Proc. Natl. Acad. Sci. U. S. A.}\ }\textbf {\bibinfo {volume}
  {(ahead of print)}},\ \bibinfo {pages} {201713538} (\bibinfo {year}
  {2017})}\BibitemShut {NoStop}%
\bibitem [{\citenamefont {Kurman}\ \emph {et~al.}(2018)\citenamefont {Kurman},
  \citenamefont {Rivera}, \citenamefont {Christensen}, \citenamefont {Tsesses},
  \citenamefont {Orenstein}, \citenamefont {Solja{\v{c}}i{\'c}}, \citenamefont
  {Joannopoulos},\ and\ \citenamefont {Kaminer}}]{kurman2018control}%
  \BibitemOpen
  \bibfield  {author} {\bibinfo {author} {\bibfnamefont {Y.}~\bibnamefont
  {Kurman}}, \bibinfo {author} {\bibfnamefont {N.}~\bibnamefont {Rivera}},
  \bibinfo {author} {\bibfnamefont {T.}~\bibnamefont {Christensen}}, \bibinfo
  {author} {\bibfnamefont {S.}~\bibnamefont {Tsesses}}, \bibinfo {author}
  {\bibfnamefont {M.}~\bibnamefont {Orenstein}}, \bibinfo {author}
  {\bibfnamefont {M.}~\bibnamefont {Solja{\v{c}}i{\'c}}}, \bibinfo {author}
  {\bibfnamefont {J.~D.}\ \bibnamefont {Joannopoulos}}, \ and\ \bibinfo
  {author} {\bibfnamefont {I.}~\bibnamefont {Kaminer}},\ }\href {\doibase
  10.1038/s41566-018-0176-6} {\bibfield  {journal} {\bibinfo  {journal} {Nat.
  Photonics}\ }\textbf {\bibinfo {volume} {12}},\ \bibinfo {pages} {423}
  (\bibinfo {year} {2018})}\BibitemShut {NoStop}%
\bibitem [{\citenamefont {Peyskens}\ and\ \citenamefont
  {Englund}(2018)}]{Peyskens:2018fk}%
  \BibitemOpen
  \bibfield  {author} {\bibinfo {author} {\bibfnamefont {F.}~\bibnamefont
  {Peyskens}}\ and\ \bibinfo {author} {\bibfnamefont {D.}~\bibnamefont
  {Englund}},\ }\href {\doibase 10.1103/PhysRevA.97.063844} {\bibfield
  {journal} {\bibinfo  {journal} {Physical Review A}\ }\textbf {\bibinfo
  {volume} {97}},\ \bibinfo {pages} {063844} (\bibinfo {year}
  {2018})}\BibitemShut {NoStop}%
\bibitem [{\citenamefont {Alcaraz~Iranzo}\ \emph {et~al.}(2018)\citenamefont
  {Alcaraz~Iranzo}, \citenamefont {Nanot}, \citenamefont {Dias}, \citenamefont
  {Epstein}, \citenamefont {Peng}, \citenamefont {Efetov}, \citenamefont
  {Lundeberg}, \citenamefont {Parret}, \citenamefont {Osmond}, \citenamefont
  {Hong}, \citenamefont {Kong}, \citenamefont {Englund}, \citenamefont
  {Peres},\ and\ \citenamefont {Koppens}}]{Alcaraz-Iranzo:2018uq}%
  \BibitemOpen
  \bibfield  {author} {\bibinfo {author} {\bibfnamefont {D.}~\bibnamefont
  {Alcaraz~Iranzo}}, \bibinfo {author} {\bibfnamefont {S.}~\bibnamefont
  {Nanot}}, \bibinfo {author} {\bibfnamefont {E.~J.~C.}\ \bibnamefont {Dias}},
  \bibinfo {author} {\bibfnamefont {I.}~\bibnamefont {Epstein}}, \bibinfo
  {author} {\bibfnamefont {C.}~\bibnamefont {Peng}}, \bibinfo {author}
  {\bibfnamefont {D.~K.}\ \bibnamefont {Efetov}}, \bibinfo {author}
  {\bibfnamefont {M.~B.}\ \bibnamefont {Lundeberg}}, \bibinfo {author}
  {\bibfnamefont {R.}~\bibnamefont {Parret}}, \bibinfo {author} {\bibfnamefont
  {J.}~\bibnamefont {Osmond}}, \bibinfo {author} {\bibfnamefont {J.-Y.}\
  \bibnamefont {Hong}}, \bibinfo {author} {\bibfnamefont {J.}~\bibnamefont
  {Kong}}, \bibinfo {author} {\bibfnamefont {D.~R.}\ \bibnamefont {Englund}},
  \bibinfo {author} {\bibfnamefont {N.~M.~R.}\ \bibnamefont {Peres}}, \ and\
  \bibinfo {author} {\bibfnamefont {F.~H.~L.}\ \bibnamefont {Koppens}},\ }\href
  {http://science.sciencemag.org/content/360/6386/291.abstract} {\bibfield
  {journal} {\bibinfo  {journal} {Science}\ }\textbf {\bibinfo {volume}
  {360}},\ \bibinfo {pages} {291} (\bibinfo {year} {2018})}\BibitemShut
  {NoStop}%
\bibitem [{\citenamefont {Jablan}\ \emph {et~al.}(2009)\citenamefont {Jablan},
  \citenamefont {Buljan},\ and\ \citenamefont
  {Solja{\v{c}}i{\'c}}}]{jablan2009plasmonics}%
  \BibitemOpen
  \bibfield  {author} {\bibinfo {author} {\bibfnamefont {M.}~\bibnamefont
  {Jablan}}, \bibinfo {author} {\bibfnamefont {H.}~\bibnamefont {Buljan}}, \
  and\ \bibinfo {author} {\bibfnamefont {M.}~\bibnamefont
  {Solja{\v{c}}i{\'c}}},\ }\href {\doibase 10.1103/PhysRevB.80.245435}
  {\bibfield  {journal} {\bibinfo  {journal} {Phys. Rev. B}\ }\textbf {\bibinfo
  {volume} {80}},\ \bibinfo {pages} {245435} (\bibinfo {year}
  {2009})}\BibitemShut {NoStop}%
\bibitem [{\citenamefont {Koppens}\ \emph {et~al.}(2011)\citenamefont
  {Koppens}, \citenamefont {Chang},\ and\ \citenamefont {Garcia~de
  Abajo}}]{koppens2011graphene}%
  \BibitemOpen
  \bibfield  {author} {\bibinfo {author} {\bibfnamefont {F.~H.~L.}\
  \bibnamefont {Koppens}}, \bibinfo {author} {\bibfnamefont {D.~E.}\
  \bibnamefont {Chang}}, \ and\ \bibinfo {author} {\bibfnamefont {F.~J.}\
  \bibnamefont {Garcia~de Abajo}},\ }\href {\doibase 10.1021/nl201771h}
  {\bibfield  {journal} {\bibinfo  {journal} {Nano Lett.}\ }\textbf {\bibinfo
  {volume} {11}},\ \bibinfo {pages} {3370} (\bibinfo {year}
  {2011})}\BibitemShut {NoStop}%
\bibitem [{\citenamefont {Fei}\ \emph {et~al.}(2012)\citenamefont {Fei} \emph
  {et~al.}}]{fei2012gate}%
  \BibitemOpen
  \bibfield  {author} {\bibinfo {author} {\bibfnamefont {Z.}~\bibnamefont
  {Fei}} \emph {et~al.},\ }\href {\doibase 10.1038/nature11253} {\bibfield
  {journal} {\bibinfo  {journal} {Nature}\ }\textbf {\bibinfo {volume} {487}},\
  \bibinfo {pages} {82} (\bibinfo {year} {2012})}\BibitemShut {NoStop}%
\bibitem [{\citenamefont {Chen}\ \emph {et~al.}(2012)\citenamefont {Chen} \emph
  {et~al.}}]{chen2012optical}%
  \BibitemOpen
  \bibfield  {author} {\bibinfo {author} {\bibfnamefont {J.}~\bibnamefont
  {Chen}} \emph {et~al.},\ }\href {\doibase 10.1038/nature11254} {\bibfield
  {journal} {\bibinfo  {journal} {Nature}\ }\textbf {\bibinfo {volume} {487}},\
  \bibinfo {pages} {77} (\bibinfo {year} {2012})}\BibitemShut {NoStop}%
\bibitem [{\citenamefont {Rivera}\ \emph {et~al.}(2016)\citenamefont {Rivera},
  \citenamefont {Kaminer}, \citenamefont {Zhen}, \citenamefont {Joannopoulos},\
  and\ \citenamefont {Solja{\v{c}}i{\'c}}}]{rivera2016shrinking}%
  \BibitemOpen
  \bibfield  {author} {\bibinfo {author} {\bibfnamefont {N.}~\bibnamefont
  {Rivera}}, \bibinfo {author} {\bibfnamefont {I.}~\bibnamefont {Kaminer}},
  \bibinfo {author} {\bibfnamefont {B.}~\bibnamefont {Zhen}}, \bibinfo {author}
  {\bibfnamefont {J.~D.}\ \bibnamefont {Joannopoulos}}, \ and\ \bibinfo
  {author} {\bibfnamefont {M.}~\bibnamefont {Solja{\v{c}}i{\'c}}},\ }\href
  {\doibase 10.1126/science.aaf6308} {\bibfield  {journal} {\bibinfo  {journal}
  {Science}\ }\textbf {\bibinfo {volume} {353}},\ \bibinfo {pages} {263}
  (\bibinfo {year} {2016})}\BibitemShut {NoStop}%
\bibitem [{\citenamefont {Lundeberg}\ \emph {et~al.}(2017)\citenamefont
  {Lundeberg} \emph {et~al.}}]{lundeberg2017tuning}%
  \BibitemOpen
  \bibfield  {author} {\bibinfo {author} {\bibfnamefont {M.~B.}\ \bibnamefont
  {Lundeberg}} \emph {et~al.},\ }\href {\doibase 10.1126/science.aan2735}
  {\bibfield  {journal} {\bibinfo  {journal} {Science}\ }\textbf {\bibinfo
  {volume} {\!}},\ \bibinfo {pages} {eaan2735} (\bibinfo {year}
  {2017})}\BibitemShut {NoStop}%
\bibitem [{\citenamefont {Zhu}\ \emph {et~al.}(2018)\citenamefont {Zhu},
  \citenamefont {Yi}, \citenamefont {Li}, \citenamefont {Xiao}, \citenamefont
  {Zhang}, \citenamefont {Yang}, \citenamefont {Kaindl}, \citenamefont {Li},
  \citenamefont {Wang},\ and\ \citenamefont {Zhang}}]{zhu2018observation}%
  \BibitemOpen
  \bibfield  {author} {\bibinfo {author} {\bibfnamefont {H.}~\bibnamefont
  {Zhu}}, \bibinfo {author} {\bibfnamefont {J.}~\bibnamefont {Yi}}, \bibinfo
  {author} {\bibfnamefont {M.-Y.}\ \bibnamefont {Li}}, \bibinfo {author}
  {\bibfnamefont {J.}~\bibnamefont {Xiao}}, \bibinfo {author} {\bibfnamefont
  {L.}~\bibnamefont {Zhang}}, \bibinfo {author} {\bibfnamefont {C.-W.}\
  \bibnamefont {Yang}}, \bibinfo {author} {\bibfnamefont {R.~A.}\ \bibnamefont
  {Kaindl}}, \bibinfo {author} {\bibfnamefont {L.-J.}\ \bibnamefont {Li}},
  \bibinfo {author} {\bibfnamefont {Y.}~\bibnamefont {Wang}}, \ and\ \bibinfo
  {author} {\bibfnamefont {X.}~\bibnamefont {Zhang}},\ }\href {\doibase
  10.1126/science.aar2711} {\bibfield  {journal} {\bibinfo  {journal}
  {Science}\ }\textbf {\bibinfo {volume} {359}},\ \bibinfo {pages} {579}
  (\bibinfo {year} {2018})}\BibitemShut {NoStop}%
\bibitem [{\citenamefont {Ratchford}\ \emph {et~al.}(2018)\citenamefont
  {Ratchford} \emph {et~al.}}]{ratchford2018controlling}%
  \BibitemOpen
  \bibfield  {author} {\bibinfo {author} {\bibfnamefont {D.~C.}\ \bibnamefont
  {Ratchford}} \emph {et~al.},\ }\href {https://arxiv.org/abs/1806.06792}
  {\bibfield  {journal} {\bibinfo  {journal} {arXiv:1806.06792}\ } (\bibinfo
  {year} {2018})}\BibitemShut {NoStop}%
\bibitem [{\citenamefont {Gonze}\ and\ \citenamefont
  {Lee}(1997)}]{Gonze1997dynamical}%
  \BibitemOpen
  \bibfield  {author} {\bibinfo {author} {\bibfnamefont {X.}~\bibnamefont
  {Gonze}}\ and\ \bibinfo {author} {\bibfnamefont {C.}~\bibnamefont {Lee}},\
  }\href {\doibase 10.1103/PhysRevB.55.10355} {\bibfield  {journal} {\bibinfo
  {journal} {Phys. Rev. B}\ }\textbf {\bibinfo {volume} {55}},\ \bibinfo
  {pages} {10355} (\bibinfo {year} {1997})}\BibitemShut {NoStop}%
\bibitem [{\citenamefont {Gonze}\ \emph {et~al.}(2009)\citenamefont {Gonze}
  \emph {et~al.}}]{abinit1}%
  \BibitemOpen
  \bibfield  {author} {\bibinfo {author} {\bibfnamefont {X.}~\bibnamefont
  {Gonze}} \emph {et~al.},\ }\href {\doibase 10.1016/j.cpc.2009.07.007}
  {\bibfield  {journal} {\bibinfo  {journal} {Comput. Phys. Commun.}\ }\textbf
  {\bibinfo {volume} {180}},\ \bibinfo {pages} {2582 } (\bibinfo {year}
  {2009})}\BibitemShut {NoStop}%
\bibitem [{\citenamefont {Gonze}\ \emph {et~al.}(2005)\citenamefont {Gonze}
  \emph {et~al.}}]{abinit2}%
  \BibitemOpen
  \bibfield  {author} {\bibinfo {author} {\bibfnamefont {X.}~\bibnamefont
  {Gonze}} \emph {et~al.},\ }\href {\doibase 10.1524/zkri.220.5.558.65066}
  {\bibfield  {journal} {\bibinfo  {journal} {Z. Kristallogr.}\ }\textbf
  {\bibinfo {volume} {220}},\ \bibinfo {pages} {558} (\bibinfo {year}
  {2005})}\BibitemShut {NoStop}%
\bibitem [{\citenamefont {Gonze}\ \emph {et~al.}(2002)\citenamefont {Gonze}
  \emph {et~al.}}]{abinit3}%
  \BibitemOpen
  \bibfield  {author} {\bibinfo {author} {\bibfnamefont {X.}~\bibnamefont
  {Gonze}} \emph {et~al.},\ }\href {\doibase 10.1016/S0927-0256(02)00325-7}
  {\bibfield  {journal} {\bibinfo  {journal} {Comput. Mater. Sci.}\ }\textbf
  {\bibinfo {volume} {25}},\ \bibinfo {pages} {478} (\bibinfo {year}
  {2002})}\BibitemShut {NoStop}%
\bibitem [{\citenamefont {Giuliani}\ and\ \citenamefont
  {Vignale}(2005)}]{giulianivignale}%
  \BibitemOpen
  \bibfield  {author} {\bibinfo {author} {\bibfnamefont {G.~F.}\ \bibnamefont
  {Giuliani}}\ and\ \bibinfo {author} {\bibfnamefont {G.}~\bibnamefont
  {Vignale}},\ }\href@noop {} {\emph {\bibinfo {title} {Quantum Theory of the
  Electron Liquid}}}\ (\bibinfo  {publisher} {Cambridge University Press},\
  \bibinfo {year} {2005})\BibitemShut {NoStop}%
\bibitem [{\citenamefont {Born}\ and\ \citenamefont
  {Huang}(1954)}]{BornHuang:1954}%
  \BibitemOpen
  \bibfield  {author} {\bibinfo {author} {\bibfnamefont {M.}~\bibnamefont
  {Born}}\ and\ \bibinfo {author} {\bibfnamefont {K.}~\bibnamefont {Huang}},\
  }\href@noop {} {\emph {\bibinfo {title} {Dynamical Theory of Crystal
  Lattices}}}\ (\bibinfo  {publisher} {Clarendon Press},\ \bibinfo {year}
  {1954})\BibitemShut {NoStop}%
\bibitem [{\citenamefont {Tiwald}\ \emph {et~al.}(1999)\citenamefont {Tiwald},
  \citenamefont {Woollam}, \citenamefont {Zollner}, \citenamefont
  {Christiansen}, \citenamefont {Gregory}, \citenamefont {Wetteroth},
  \citenamefont {Wilson},\ and\ \citenamefont {Powell}}]{tiwald1999carrier}%
  \BibitemOpen
  \bibfield  {author} {\bibinfo {author} {\bibfnamefont {T.~E.}\ \bibnamefont
  {Tiwald}}, \bibinfo {author} {\bibfnamefont {J.~A.}\ \bibnamefont {Woollam}},
  \bibinfo {author} {\bibfnamefont {S.}~\bibnamefont {Zollner}}, \bibinfo
  {author} {\bibfnamefont {J.}~\bibnamefont {Christiansen}}, \bibinfo {author}
  {\bibfnamefont {R.}~\bibnamefont {Gregory}}, \bibinfo {author} {\bibfnamefont
  {T.}~\bibnamefont {Wetteroth}}, \bibinfo {author} {\bibfnamefont
  {S.}~\bibnamefont {Wilson}}, \ and\ \bibinfo {author} {\bibfnamefont {A.~R.}\
  \bibnamefont {Powell}},\ }\href {\doibase 10.1103/PhysRevB.60.11464}
  {\bibfield  {journal} {\bibinfo  {journal} {Phys. Rev. B}\ }\textbf {\bibinfo
  {volume} {60}},\ \bibinfo {pages} {11464} (\bibinfo {year}
  {1999})}\BibitemShut {NoStop}%
\bibitem [{\citenamefont {Iranzo}\ \emph {et~al.}(2018)\citenamefont {Iranzo}
  \emph {et~al.}}]{iranzo2018probing}%
  \BibitemOpen
  \bibfield  {author} {\bibinfo {author} {\bibfnamefont {D.~A.}\ \bibnamefont
  {Iranzo}} \emph {et~al.},\ }\href {\doibase 10.1126/science.aar8438}
  {\bibfield  {journal} {\bibinfo  {journal} {Science}\ }\textbf {\bibinfo
  {volume} {360}},\ \bibinfo {pages} {291} (\bibinfo {year}
  {2018})}\BibitemShut {NoStop}%
\bibitem [{\citenamefont {Benz}\ \emph {et~al.}(2016)\citenamefont {Benz} \emph
  {et~al.}}]{benz2016single}%
  \BibitemOpen
  \bibfield  {author} {\bibinfo {author} {\bibfnamefont {F.}~\bibnamefont
  {Benz}} \emph {et~al.},\ }\href {\doibase 10.1126/science.aah5243} {\bibfield
   {journal} {\bibinfo  {journal} {Science}\ }\textbf {\bibinfo {volume}
  {354}},\ \bibinfo {pages} {726} (\bibinfo {year} {2016})}\BibitemShut
  {NoStop}%
\bibitem [{\citenamefont {Hamann}\ \emph {et~al.}(2005)\citenamefont {Hamann},
  \citenamefont {Wu}, \citenamefont {Rabe},\ and\ \citenamefont
  {Vanderbilt}}]{Hamann2005metric}%
  \BibitemOpen
  \bibfield  {author} {\bibinfo {author} {\bibfnamefont {D.}~\bibnamefont
  {Hamann}}, \bibinfo {author} {\bibfnamefont {X.}~\bibnamefont {Wu}}, \bibinfo
  {author} {\bibfnamefont {K.~M.}\ \bibnamefont {Rabe}}, \ and\ \bibinfo
  {author} {\bibfnamefont {D.}~\bibnamefont {Vanderbilt}},\ }\href {\doibase
  10.1103/PhysRevB.71.035117} {\bibfield  {journal} {\bibinfo  {journal} {Phys.
  Rev. B}\ }\textbf {\bibinfo {volume} {71}},\ \bibinfo {pages} {035117}
  (\bibinfo {year} {2005})}\BibitemShut {NoStop}%
\bibitem [{\citenamefont {Perdew}\ \emph {et~al.}(2008)\citenamefont {Perdew},
  \citenamefont {Ruzsinszky}, \citenamefont {Csonka}, \citenamefont {Vydrov},
  \citenamefont {Scuseria}, \citenamefont {Constantin}, \citenamefont {Zhou},\
  and\ \citenamefont {Burke}}]{PBEsol}%
  \BibitemOpen
  \bibfield  {author} {\bibinfo {author} {\bibfnamefont {J.~P.}\ \bibnamefont
  {Perdew}}, \bibinfo {author} {\bibfnamefont {A.}~\bibnamefont {Ruzsinszky}},
  \bibinfo {author} {\bibfnamefont {G.~I.}\ \bibnamefont {Csonka}}, \bibinfo
  {author} {\bibfnamefont {O.~A.}\ \bibnamefont {Vydrov}}, \bibinfo {author}
  {\bibfnamefont {G.~E.}\ \bibnamefont {Scuseria}}, \bibinfo {author}
  {\bibfnamefont {L.~A.}\ \bibnamefont {Constantin}}, \bibinfo {author}
  {\bibfnamefont {X.}~\bibnamefont {Zhou}}, \ and\ \bibinfo {author}
  {\bibfnamefont {K.}~\bibnamefont {Burke}},\ }\href {\doibase
  10.1103/PhysRevLett.100.136406} {\bibfield  {journal} {\bibinfo  {journal}
  {Phys. Rev. Lett.}\ }\textbf {\bibinfo {volume} {100}},\ \bibinfo {pages}
  {136406} (\bibinfo {year} {2008})}\BibitemShut {NoStop}%
\bibitem [{\citenamefont {Perdew}\ \emph {et~al.}(1996)\citenamefont {Perdew},
  \citenamefont {Burke},\ and\ \citenamefont {Ernzerhof}}]{PBE}%
  \BibitemOpen
  \bibfield  {author} {\bibinfo {author} {\bibfnamefont {J.~P.}\ \bibnamefont
  {Perdew}}, \bibinfo {author} {\bibfnamefont {K.}~\bibnamefont {Burke}}, \
  and\ \bibinfo {author} {\bibfnamefont {M.}~\bibnamefont {Ernzerhof}},\ }\href
  {\doibase 10.1103/PhysRevLett.77.3865} {\bibfield  {journal} {\bibinfo
  {journal} {Phys. Rev. Lett.}\ }\textbf {\bibinfo {volume} {77}},\ \bibinfo
  {pages} {3865} (\bibinfo {year} {1996})}\BibitemShut {NoStop}%
\bibitem [{\citenamefont {Hamann}(2013)}]{ONCV_PPs}%
  \BibitemOpen
  \bibfield  {author} {\bibinfo {author} {\bibfnamefont {D.~R.}\ \bibnamefont
  {Hamann}},\ }\href {\doibase 10.1103/PhysRevB.88.085117} {\bibfield
  {journal} {\bibinfo  {journal} {Phys. Rev. B}\ }\textbf {\bibinfo {volume}
  {88}},\ \bibinfo {pages} {085117} (\bibinfo {year} {2013})}\BibitemShut
  {NoStop}%
\bibitem [{\citenamefont {van Setten}\ \emph {et~al.}(2018)\citenamefont {van
  Setten}, \citenamefont {Giantomassi}, \citenamefont {Bousquet}, \citenamefont
  {Verstraete}, \citenamefont {Hamann}, \citenamefont {Gonze},\ and\
  \citenamefont {Rignanese}}]{pseudodojo}%
  \BibitemOpen
  \bibfield  {author} {\bibinfo {author} {\bibfnamefont {M.}~\bibnamefont {van
  Setten}}, \bibinfo {author} {\bibfnamefont {M.}~\bibnamefont {Giantomassi}},
  \bibinfo {author} {\bibfnamefont {E.}~\bibnamefont {Bousquet}}, \bibinfo
  {author} {\bibfnamefont {M.}~\bibnamefont {Verstraete}}, \bibinfo {author}
  {\bibfnamefont {D.}~\bibnamefont {Hamann}}, \bibinfo {author} {\bibfnamefont
  {X.}~\bibnamefont {Gonze}}, \ and\ \bibinfo {author} {\bibfnamefont {G.-M.}\
  \bibnamefont {Rignanese}},\ }\href {\doibase
  https://doi.org/10.1016/j.cpc.2018.01.012} {\bibfield  {journal} {\bibinfo
  {journal} {Comput. Phys. Commun.}\ }\textbf {\bibinfo {volume} {226}},\
  \bibinfo {pages} {39 } (\bibinfo {year} {2018})}\BibitemShut {NoStop}%
\bibitem [{\citenamefont {Grimme}\ \emph {et~al.}(2010)\citenamefont {Grimme},
  \citenamefont {Antony}, \citenamefont {Ehrlich},\ and\ \citenamefont
  {Krieg}}]{grimme}%
  \BibitemOpen
  \bibfield  {author} {\bibinfo {author} {\bibfnamefont {S.}~\bibnamefont
  {Grimme}}, \bibinfo {author} {\bibfnamefont {J.}~\bibnamefont {Antony}},
  \bibinfo {author} {\bibfnamefont {S.}~\bibnamefont {Ehrlich}}, \ and\
  \bibinfo {author} {\bibfnamefont {H.}~\bibnamefont {Krieg}},\ }\href
  {\doibase 10.1063/1.3382344} {\bibfield  {journal} {\bibinfo  {journal} {J.
  Chem. Phys.}\ }\textbf {\bibinfo {volume} {132}},\ \bibinfo {pages} {154104}
  (\bibinfo {year} {2010})}\BibitemShut {NoStop}%
\bibitem [{\citenamefont {Ahuja}(2002)}]{band_gap_expt1}%
  \BibitemOpen
  \bibfield  {author} {\bibinfo {author} {\bibfnamefont {R.}~\bibnamefont
  {Ahuja}},\ }\href {\doibase 10.1063/1.1429766} {\bibfield  {journal}
  {\bibinfo  {journal} {J. Appl. Phys.}\ }\textbf {\bibinfo {volume} {91}},\
  \bibinfo {pages} {2099} (\bibinfo {year} {2002})}\BibitemShut {NoStop}%
\bibitem [{\citenamefont {Haberstroh}\ \emph {et~al.}(1994)\citenamefont
  {Haberstroh}, \citenamefont {Helbig},\ and\ \citenamefont
  {Stein}}]{band_gap_expt2}%
  \BibitemOpen
  \bibfield  {author} {\bibinfo {author} {\bibfnamefont {{\relax
  Ch.}.}~\bibnamefont {Haberstroh}}, \bibinfo {author} {\bibfnamefont
  {R.}~\bibnamefont {Helbig}}, \ and\ \bibinfo {author} {\bibfnamefont {R.~A.}\
  \bibnamefont {Stein}},\ }\href {\doibase 10.1063/1.357103} {\bibfield
  {journal} {\bibinfo  {journal} {J. Appl. Phys.}\ }\textbf {\bibinfo {volume}
  {76}},\ \bibinfo {pages} {509} (\bibinfo {year} {1994})}\BibitemShut
  {NoStop}%
\bibitem [{\citenamefont {Sundararaman}\ \emph {et~al.}(2014)\citenamefont
  {Sundararaman}, \citenamefont {Narang}, \citenamefont {Jermyn}, \citenamefont
  {Goddard~III},\ and\ \citenamefont {Atwater}}]{sundararaman2014theoretical}%
  \BibitemOpen
  \bibfield  {author} {\bibinfo {author} {\bibfnamefont {R.}~\bibnamefont
  {Sundararaman}}, \bibinfo {author} {\bibfnamefont {P.}~\bibnamefont
  {Narang}}, \bibinfo {author} {\bibfnamefont {A.~S.}\ \bibnamefont {Jermyn}},
  \bibinfo {author} {\bibfnamefont {W.~A.}\ \bibnamefont {Goddard~III}}, \ and\
  \bibinfo {author} {\bibfnamefont {H.~A.}\ \bibnamefont {Atwater}},\ }\href
  {\doibase 10.1038/ncomms6788} {\bibfield  {journal} {\bibinfo  {journal}
  {Nat. Commun.}\ }\textbf {\bibinfo {volume} {5}},\ \bibinfo {pages} {5788}
  (\bibinfo {year} {2014})}\BibitemShut {NoStop}%
\bibitem [{\citenamefont {Brown}\ \emph {et~al.}(2015)\citenamefont {Brown},
  \citenamefont {Sundararaman}, \citenamefont {Narang}, \citenamefont
  {Goddard~III},\ and\ \citenamefont {Atwater}}]{brown2015nonradiative}%
  \BibitemOpen
  \bibfield  {author} {\bibinfo {author} {\bibfnamefont {A.~M.}\ \bibnamefont
  {Brown}}, \bibinfo {author} {\bibfnamefont {R.}~\bibnamefont {Sundararaman}},
  \bibinfo {author} {\bibfnamefont {P.}~\bibnamefont {Narang}}, \bibinfo
  {author} {\bibfnamefont {W.~A.}\ \bibnamefont {Goddard~III}}, \ and\ \bibinfo
  {author} {\bibfnamefont {H.~A.}\ \bibnamefont {Atwater}},\ }\href {\doibase
  10.1021/acsnano.5b06199} {\bibfield  {journal} {\bibinfo  {journal} {ACS
  Nano}\ }\textbf {\bibinfo {volume} {10}},\ \bibinfo {pages} {957} (\bibinfo
  {year} {2015})}\BibitemShut {NoStop}%
\bibitem [{\citenamefont {Brown}\ \emph {et~al.}(2016)\citenamefont {Brown},
  \citenamefont {Sundararaman}, \citenamefont {Narang}, \citenamefont
  {Goddard~III},\ and\ \citenamefont {Atwater}}]{brown2016ab}%
  \BibitemOpen
  \bibfield  {author} {\bibinfo {author} {\bibfnamefont {A.~M.}\ \bibnamefont
  {Brown}}, \bibinfo {author} {\bibfnamefont {R.}~\bibnamefont {Sundararaman}},
  \bibinfo {author} {\bibfnamefont {P.}~\bibnamefont {Narang}}, \bibinfo
  {author} {\bibfnamefont {W.~A.}\ \bibnamefont {Goddard~III}}, \ and\ \bibinfo
  {author} {\bibfnamefont {H.~A.}\ \bibnamefont {Atwater}},\ }\href {\doibase
  10.1103/PhysRevB.94.075120} {\bibfield  {journal} {\bibinfo  {journal} {Phys.
  Rev. B}\ }\textbf {\bibinfo {volume} {94}},\ \bibinfo {pages} {075120}
  (\bibinfo {year} {2016})}\BibitemShut {NoStop}%
\bibitem [{\citenamefont {Narang}\ \emph {et~al.}(2017)\citenamefont {Narang},
  \citenamefont {Zhao}, \citenamefont {Claybrook},\ and\ \citenamefont
  {Sundararaman}}]{narang2017effects}%
  \BibitemOpen
  \bibfield  {author} {\bibinfo {author} {\bibfnamefont {P.}~\bibnamefont
  {Narang}}, \bibinfo {author} {\bibfnamefont {L.}~\bibnamefont {Zhao}},
  \bibinfo {author} {\bibfnamefont {S.}~\bibnamefont {Claybrook}}, \ and\
  \bibinfo {author} {\bibfnamefont {R.}~\bibnamefont {Sundararaman}},\ }\href
  {\doibase 10.1002/adom.201600914} {\bibfield  {journal} {\bibinfo  {journal}
  {Adv. Opt. Mater.}\ }\textbf {\bibinfo {volume} {5}},\ \bibinfo {pages}
  {1600914} (\bibinfo {year} {2017})}\BibitemShut {NoStop}%
\bibitem [{\citenamefont {Srivastava}(1990)}]{srivastava1990physics}%
  \BibitemOpen
  \bibfield  {author} {\bibinfo {author} {\bibfnamefont {G.~P.}\ \bibnamefont
  {Srivastava}},\ }\href@noop {} {\emph {\bibinfo {title} {The Physics of
  Phonons}}}\ (\bibinfo  {publisher} {CRC Press},\ \bibinfo {year}
  {1990})\BibitemShut {NoStop}%
\bibitem [{\citenamefont {Robert}\ \emph {et~al.}(2016)\citenamefont {Robert}
  \emph {et~al.}}]{robert2016excitonic}%
  \BibitemOpen
  \bibfield  {author} {\bibinfo {author} {\bibfnamefont {C.}~\bibnamefont
  {Robert}} \emph {et~al.},\ }\href {\doibase 10.1103/PhysRevB.94.155425}
  {\bibfield  {journal} {\bibinfo  {journal} {Phys. Rev. B}\ }\textbf {\bibinfo
  {volume} {94}},\ \bibinfo {pages} {155425} (\bibinfo {year}
  {2016})}\BibitemShut {NoStop}%
\bibitem [{\citenamefont {Levinshtein}\ \emph {et~al.}(2001)\citenamefont
  {Levinshtein}, \citenamefont {Rumyantsev},\ and\ \citenamefont
  {Shur}}]{levinshtein2001properties}%
  \BibitemOpen
  \bibfield  {author} {\bibinfo {author} {\bibfnamefont {M.~E.}\ \bibnamefont
  {Levinshtein}}, \bibinfo {author} {\bibfnamefont {S.~L.}\ \bibnamefont
  {Rumyantsev}}, \ and\ \bibinfo {author} {\bibfnamefont {M.~S.}\ \bibnamefont
  {Shur}},\ }\href@noop {} {\emph {\bibinfo {title} {Properties of Advanced
  Semiconductor Materials: GaN, AIN, InN, BN, SiC, SiGe}}}\ (\bibinfo
  {publisher} {John Wiley \& Sons},\ \bibinfo {year} {2001})\BibitemShut
  {NoStop}%
\bibitem [{\citenamefont {Attaccalite}\ \emph {et~al.}(2018)\citenamefont
  {Attaccalite}, \citenamefont {Gr{\"u}ning}, \citenamefont {Amara},
  \citenamefont {Latil},\ and\ \citenamefont
  {Ducastelle}}]{attaccalite2018two}%
  \BibitemOpen
  \bibfield  {author} {\bibinfo {author} {\bibfnamefont {C.}~\bibnamefont
  {Attaccalite}}, \bibinfo {author} {\bibfnamefont {M.}~\bibnamefont
  {Gr{\"u}ning}}, \bibinfo {author} {\bibfnamefont {H.}~\bibnamefont {Amara}},
  \bibinfo {author} {\bibfnamefont {S.}~\bibnamefont {Latil}}, \ and\ \bibinfo
  {author} {\bibfnamefont {F.}~\bibnamefont {Ducastelle}},\ }\href
  {https://arxiv.org/abs/1803.10959} {\bibfield  {journal} {\bibinfo  {journal}
  {arXiv:1803.10959}\ } (\bibinfo {year} {2018})}\BibitemShut {NoStop}%
\bibitem [{\citenamefont {Manjavacas}\ \emph {et~al.}(2013)\citenamefont
  {Manjavacas}, \citenamefont {Marchesin}, \citenamefont {Thongrattanasiri},
  \citenamefont {Koval}, \citenamefont {Nordlander}, \citenamefont
  {S{\'a}nchez-Portal},\ and\ \citenamefont {Garc{\'\i}a~de
  Abajo}}]{manjavacas2013tunable}%
  \BibitemOpen
  \bibfield  {author} {\bibinfo {author} {\bibfnamefont {A.}~\bibnamefont
  {Manjavacas}}, \bibinfo {author} {\bibfnamefont {F.}~\bibnamefont
  {Marchesin}}, \bibinfo {author} {\bibfnamefont {S.}~\bibnamefont
  {Thongrattanasiri}}, \bibinfo {author} {\bibfnamefont {P.}~\bibnamefont
  {Koval}}, \bibinfo {author} {\bibfnamefont {P.}~\bibnamefont {Nordlander}},
  \bibinfo {author} {\bibfnamefont {D.}~\bibnamefont {S{\'a}nchez-Portal}}, \
  and\ \bibinfo {author} {\bibfnamefont {F.~J.}\ \bibnamefont {Garc{\'\i}a~de
  Abajo}},\ }\href {\doibase 10.1021/nn4006297} {\bibfield  {journal} {\bibinfo
   {journal} {ACS Nano}\ }\textbf {\bibinfo {volume} {7}},\ \bibinfo {pages}
  {3635} (\bibinfo {year} {2013})}\BibitemShut {NoStop}%
\bibitem [{\citenamefont {Lauchner}\ \emph {et~al.}(2015)\citenamefont
  {Lauchner}, \citenamefont {Schlather}, \citenamefont {Manjavacas},
  \citenamefont {Cui}, \citenamefont {McClain}, \citenamefont {Stec},
  \citenamefont {Garc{\'\i}a~de Abajo}, \citenamefont {Nordlander},\ and\
  \citenamefont {Halas}}]{lauchner2015molecular}%
  \BibitemOpen
  \bibfield  {author} {\bibinfo {author} {\bibfnamefont {A.}~\bibnamefont
  {Lauchner}}, \bibinfo {author} {\bibfnamefont {A.~E.}\ \bibnamefont
  {Schlather}}, \bibinfo {author} {\bibfnamefont {A.}~\bibnamefont
  {Manjavacas}}, \bibinfo {author} {\bibfnamefont {Y.}~\bibnamefont {Cui}},
  \bibinfo {author} {\bibfnamefont {M.~J.}\ \bibnamefont {McClain}}, \bibinfo
  {author} {\bibfnamefont {G.~J.}\ \bibnamefont {Stec}}, \bibinfo {author}
  {\bibfnamefont {F.~J.}\ \bibnamefont {Garc{\'\i}a~de Abajo}}, \bibinfo
  {author} {\bibfnamefont {P.}~\bibnamefont {Nordlander}}, \ and\ \bibinfo
  {author} {\bibfnamefont {N.~J.}\ \bibnamefont {Halas}},\ }\href {\doibase
  10.1021/acs.nanolett.5b02549} {\bibfield  {journal} {\bibinfo  {journal}
  {Nano Lett.}\ }\textbf {\bibinfo {volume} {15}},\ \bibinfo {pages} {6208}
  (\bibinfo {year} {2015})}\BibitemShut {NoStop}%
\end{thebibliography}%



\end{document}